\input{aipcheck}

\documentclass[
    ,final            
  ]
  {aipproc}

\layoutstyle{8x11single}

\newcommand{\nco}{\newcommand}
\nco{\beq}{\begin{equation}} \nco{\eeq}{\end{equation}}
\nco{\beqa}{\begin{eqnarray}} \nco{\eeqa}{\end{eqnarray}}
\nco{\lsim}{\mbox{\raisebox{-.6ex}{~$\stackrel{<}{\sim}$~}}}
\nco{\gsim}{\mbox{\raisebox{-.6ex}{~$\stackrel{>}{\sim}$~}}}

\def\exd{{\rm d}}

\begin{document}

\title{Towards a Natural Theory of Dark Energy:\\
Supersymmetric Large Extra Dimensions\footnote{To appear in the
proceedings of the Texas A\&M Workshop on String Cosmology.}}

\author{C.P. Burgess}{
  address={Physics Department, McGill University,
  Montr\'eal, Qu\'ebec, Canada.\\[2mm]
  Department of Physics and Astronomy, McMaster University,\\
  Hamilton, Ontario, Canada.\\[2mm]
  Perimeter Institute, Waterloo, Ontario, Canada.} }

\begin{abstract}
 The first part of this article summarizes the evidence for Dark Energy
 and Dark Matter, as well as the naturalness issues which plague
 current theories of Dark Energy. The main point of this part is to argue
 why these naturalness issues should provide the central theoretical
 guidance for the search for a successful theory. The second part of
 the article describes the present status of what I regard as being the best
 mechanism yet proposed for addressing this issue: Six-dimensional
 Supergravity with submillimetre-sized Extra Dimensions (Supersymmetric
 Large Extra Dimensions, or SLED for short). Besides
 summarizing the SLED proposal itself, this section also describes the
 tests which this model has passed, the main criticisms which have been
 raised, and the remaining challenges which remain to be checked.
 The bottom line is that the proposal survives the tests which
 have been completed to date, and predicts several distinctive
 experimental signatures for cosmology, tests of gravity and for
 accelerator-based particle physics.
\end{abstract}

\maketitle


\section{CONTENTS}

\begin{enumerate}
\item {\bf DARK ENERGY}
\medskip
\begin{enumerate}
\item {\bf Dark Matter and Dark Energy: The Evidence:} {\it ~~i.
Dark Matter; ~~ii. Dark Energy.}
\item {\bf Modelling Dark Energy:} {\it ~~i. Scales; ~~ii.
Sensitivity to Initial Conditions.}
\item{\bf Naturalness Issues:} {\it ~~i. The Problem of Quantum
Corrections; ~~ii. Two Categories of Fine Tuning; \\~~iii.
`Technical' Naturalness; ~~iv. The Problem of `Self Tuning':
Weinberg's No-Go Theorem.}
\end{enumerate}
\medskip
\item {\bf SUPERSYMMETRIC LARGE EXTRA DIMENSIONS}
\medskip
\begin{enumerate}
\item {\bf The Proposal:} {\it ~~i. The Action.}
\item {\bf Classical Relaxation of the 4D Cosmological Constant:}
{\it ~~i. 6D Self-Tuning: A Cartoon; ~~ii. 6D Self-Tuning I:
General Solutions; ~~iii. 6D Self-Tuning II: Response to Tension
Changes; ~~iv. 6D Self-Tuning III: A 4D Analysis?}
\item {\bf Quantum Bulk Contributions: Why There is Something and
Not Nothing:} {\it ~~i. Bulk Loops I: UV Sensitivity; ~~ii. Bulk
Loops II: Weinberg's Theorem Revisited; ~~iii. Bulk Loops III:
Massless 6D Modes; ~~iv. Embedding into More Fundamental
Theories?}
\item {\bf Observational Implications:} {\it ~~i. Bounds on
Extra-Dimensional Radii;~~ ii. Post-BBN Cosmology;\\~~ iii.
Pre-BBN Cosmology;~~iv. Tests of Gravity;~~ v. Implications for
Particle Physics.}
\end{enumerate}
\end{enumerate}

\section{Dark Energy}

The conventional picture of Big Bang Cosmology was stood on its
ear at the close of the 20th Century by observations which
established for the first time an observational basis for
identifying all of the main contributions to the energy density of
the present-day universe.

The cosmic surprise was to find that the universe is dominated by
{\it two} different kinds of unknown forms of matter
\cite{dmdereview,PDG}. Something like 25\% of the universal energy
density turns out to consist of what has come to be known as Dark
Matter -- a form of nonluminous matter whose existence had been
suspected for decades due to its gravitational influence on
galaxies and clusters of galaxies. More remarkably, a further 70\%
apparently consists of a different kind of unknown substance
called Dark Energy \cite{ccnonzero}. In what may be the ultimate
Copernican revolution, ordinary atoms ({\it i.e.} baryons) turn
out to make up {\it at most} only about 5\% of the universe's
contents!

The discovery of this Dark Energy has held the theorists' hands to
the fire, because it adds a new dimension to a long-standing
problem: the cosmological constant problem (for a review of which,
see \cite{ccreview}). This problem arises once one tries to embed
cosmology into a fundamental theory of small-distance physics, in
terms of which the cosmological constant can be interpreted as the
vacuum energy density. However given that microscopic theories all
tend to predict vacuum energies which are of order the microscopic
mass (or length) scales, their predictions are typically 50 orders
of magnitude larger than the observed value. The cosmological
constant problem then asks why the observed cosmological constant
should be so small, and for decades this problem was framed in
terms of trying to understand why it should be zero. With the
discovery of Dark Energy zero is no longer a good enough answer,
and a good theory must also explain why it is nonzero, and yet
small.

This first part of the review you are reading summarizes very
briefly the evidence for Dark Energy (and Dark Matter), as well as
the seemingly insurmountable theoretical problems which its
explanation raises. This is followed, in the second part of the
review, by a description of what I regard to be a very promising
line of approach to solving the cosmological constant problem; one
which purports to explain both why it is small {\it and} why it is
nonzero. The second part describes the progress to date on
understanding whether the proposal really works (result: so far,
so good), and also briefly describes several of the proposal's
many non-cosmological implications, such as for tests of General
Relativity on large and small scales and for experiments at
high-energy particle accelerators like the Large Hadron Collider
(LHC).

\subsection{Dark Matter and Dark Energy: The Evidence}

As might be expected, widespread acceptance of such a
revolutionary picture of the universe has required the concordance
of several independent lines of evidence. Although this evidence
is still improving, it has led to a remarkably robust picture.

\subsubsection{Dark Matter}

The evidence for Dark Matter goes back to the 1930s, and comes
from various methods for comparing the amount of matter which
gravitates with the amount of matter which is luminous, and so
directly visible. Several types of independent comparisons
consistently point to there being more than 10 times as much dark
material in space than is visible,\footnote{This is consistent
with the cosmological evidence that Dark Matter is roughly 5 times
more abundant than ordinary matter (baryons) because most of the
ordinary matter is also dark, and so is not visible.} with the
evidence coming from several sources:
\begin{itemize}
\item {\it Baryon Abundance Inferred from Nucleosynthesis:} The
total mass density of ordinary matter (baryons) which can be
present within the universe within the Hot Big Bang model can be
inferred in two separate ways, independently of any present-day
measurements of its gravitational properties. First, the total
predicted relative abundance of primordial nuclei within the Hot
Big Bang relies on the competition between nuclear reaction rates
and the rate with which the universe cools. But the reaction rates
depend on the net abundance of baryons in the universe, and the
cooling rate depends on the overall expansion rate of the
universe, and so -- according to General Relativity -- on its
total energy density. The success of the predictions of Big Bang
Nucleosynthesis (BBN) therefore fixes the fraction of the
universal energy density which can consist of baryons, and implies
that there can only be a few times more baryons out there than
what would be inferred by counting those which are luminous, and
so directly visible.
\item {\it Baryon Abundance Inferred from the Cosmic Microwave
Background (CMB):} The observation of CMB photons, which come to
us from directly from the Big Bang, provide an independent
indication of the overall baryon abundance. They do so because
these photons were emitted as the universe first became
transparent as it cooled enough for electrons and nuclei to form
neutral atoms. Any sound waves in the baryon density which were
present at this epoch are observable through the small temperature
fluctuations which they imply for this CMB radiation, and since
the properties of these sound waves depend on the density of
baryons at the time, a detailed understanding of the CMB spectrum
allows the total baryon density to be inferred, with a result
which is consistent with the results of Big Bang Nucleosynthesis.
\item {\it Galaxies:} The total mass in a galaxy may be inferred
from the observed galactic rotation speeds, measured as a function
of the distance from the galactic center. Since the galactic
rotation is due to stars and gas clouds orbiting the galactic
center, its speed measures the galactic mass distribution in much
the same way as Kepler's Laws can be used to infer the masses of
planets in our solar system given the orbital properties of their
moons. For large galaxies like the Milky Way, the results point to
there being several times more matter present than would be
obtained by simply counting what is directly visible. Although
this matter could consist of baryons without being in conflict
with the above bounds on the total baryon abundance, our knowledge
of the properties of baryons allows the inference that much of it
should be clumped as massive planet-like objects. Direct searches
for the microlensing which these objects would produce as they
pass in front of more distant stars indicates (weakly) that there
are not enough of these kinds of objects to account for the total
galactic mass.
\item {\it Clusters of Galaxies:} The comparison of the amount of
visible matter (in galaxies and hot intergalactic gas) in large
galaxy clusters with the total mass of the cluster inferred
gravitationally points to the existence of much more mass than is
visible. Moreover, the amount of mass which must be present is
also more than is permitted by the above upper limits on the total
baryon density, indicating that the Dark Matter cannot be made of
ordinary atoms. The total amount of cluster mass present is
inferred in several different ways, such as ($i$) by finding the
depth of the potential well which is required in order to
gravitationally bind the hot intergalactic gas, or ($ii$) by
observing the gravitational lensing of more distant galaxies as
their light passes by the foreground galaxy cluster.
\item {\it Structure Formation:} More evidence comes from the
assumption that present-day galaxies and galaxy clusters formed
due to the gravitational amplification of initially-small
primordial density fluctuations. In this case the evidence for
Dark Matter arises because of the interplay of two facts: First,
the amplitude of initial fluctuations is known to be very small,
$\delta \rho/\rho \sim 10^{-5}$, at the time when the universe
became transparent because they can be directly inferred from the
properties of the observed temperature fluctuations of the cosmic
microwave background (CMB) radiation. Second, because small
initial fluctuations cannot be amplified by gravity within a
radiation-dominated universe, fluctuations cannot begin to be
amplified until the epoch where the energy density of
non-relativistic matter begins to dominate the
more-quickly-falling energy density in radiation. There has not
been enough time for these initially-small density fluctuations to
form gravitational structure unless there is much more matter
present than can be accounted for by baryons. The amount required
agrees with the amount inferred from galaxy clusters and from CMB
measurements.
\end{itemize}

We are led to a coherent picture wherein the total amount of Dark
Matter is about five times larger than the total amount of baryons
which can be present (which is itself several times larger than
the amount of visible, luminous matter). Although we do not know
what this Dark Matter is, we do know that it is not baryons and
that its equation of state (pressure, $p$, as a function of energy
density, $\rho$) must permit the formation of gravitational
structure. This would be true, for instance, if it consisted of an
unknown species of particle which does not take part in ordinary
nuclear or electromagnetic interactions. If such a particle were
sufficiently massive then its pressure would be negligible
compared with its energy density: $p \approx 0$, just like for
baryons, and it would not hinder structure formation. Better yet,
if such a particle were stable and had a mass and interaction
strength with other matter which was similar to that of the $Z$
boson (which is {\it not} stable), its relic thermal abundance in
the Hot Big Bang would naturally be in the observed range
\cite{dmdereview,PDG}.

{}From a theoretical point of view, there seem to be two options
for explaining these observations. Since the existence of the Dark
Matter is inferred gravitationally, either the laws of gravity are
different on extra-galactic scales than those we know and test in
the solar system, or there exists a cosmic abundance of a new type
of hitherto-undiscovered Dark Matter.

The current consensus is that it is more likely that Dark Matter
is explained by the presence of a new type of particle than by
changing gravity on long distances (see however refs.~
\cite{MOND,DGP,PeeblesRatra2} for the best-studied such attempts).
This is because we know that it is fairly easy to change the laws
of gravity on very short distances, and so long as these changes
occur over distances shorter than $\sim 100$ $\mu$m their effects
could have escaped detection. But it is very difficult (but not
necessarily impossible) to sensibly modify gravity only at longer
distances without coming into conflict with observations, or with
fundamental principles of relativity or quantum mechanics. Despite
many searches for changes to gravity on long distances which can
account for the evidence for Dark Matter (and despite reasonable
success for galaxy rotation curves alone \cite{MOND}, no
completely successful candidate theory has yet been found.

On the other hand, the existence of matter of the type required to
be Dark Matter is actually predicted by many current theories of
fundamental physics. Such theories often imply the existence of
new particles whose interactions and mass are similar to the $Z$
boson, and some of these particles are often stable (such as is
true for the lightest supersymmetric particle in many
supersymmetric theories, for instance). As mentioned above, this
is just what is required to ensure that their present-day relic
abundance is what is found for Dark Matter. Although the discovery
of such a new stable particle would be a revolutionary
development, its properties seem relatively easy to fit into our
framework of microscopic theories.

The same seems not to be true for Dark Energy, as we shall now
see.

\subsubsection{Dark Energy}

Although the evidence for Dark Matter has been accumulating for
decades, the big surprise of recent years was the discovery of a
second type of dark component to the universal energy density: the
Dark Energy. This evidence for the existence of this component
comes from two independent lines of argument:

\begin{enumerate}
\item {\it Universal Acceleration:} Since gravity is attractive,
the generic behaviour which is predicted for an expanding universe
containing ordinary (and dark) matter ($p \sim 0$) or radiation
($p = \rho/3$) is that its expansion rate should be decelerating.
Its expansion decelerates because of the retarding influence of
the mutual gravitational attraction of all of the constituent
matter and radiation. So it was a surprise when the universal
expansion rate was measured at cosmological distances and found to
be {\it smaller} than it is at present, indicating that the
universal expansion is {\it accelerating} rather than
decelerating. This measurement was made by comparing the observed
brightness of various distant supernova, whose intrinsic
brightness is believed to be well understood (making them standard
candles). Comparing their observed brightness with their known
luminosity allows an inference of their distance, and so the
universal acceleration is obtained by also performing a
measurement of their redshift in order to get the expansion rate
at this distance. Since General Relativity implies that an
equation of state satisfying $p < - \rho/3$ would be required in
order to make the universal expansion accelerate, neither ordinary
matter or Dark Matter cannot be responsible. The matter whose
pressure is sufficiently negative to cause this accelerated
expansion must be another new form, and is called the Dark Energy.
\item {\it Flatness of the universe:} An independent measure of
the Dark Energy comes from the observed spectrum of temperature
fluctuations in the CMB. This is sensitive to the existence of
Dark Energy because the CMB photons have traversed the observable
universe before reaching us and so their arrival direction depends
on the overall geometry of the universe as a whole. However
General Relativity implies that the geometry of the universe also
depends on its total energy density, and (taken together with
independent measurements of the universal expansion rate) this can
be compared with the known (as above) density of ordinary matter
and Dark Matter. What is found is that the ordinary matter and
Dark Matter abundances fall short by an amount which is consistent
with the existence of Dark Energy in the amount required by the
Supernova measurements.
\end{enumerate}

{}From a theoretical perspective, we know that Dark Energy is
different from Dark Matter because its equation of state must
satisfy $p < -\rho/3$ (while Dark Matter satisfies $p \sim 0$).
Interestingly enough, this kind of stress energy {\it can} be
arranged within relativistic field theories, with the simplest
example being given by the spatially-homogeneous motion of a
scalar field (or fields), $\phi^i(t)$ \cite{PeeblesRatra1}. For
such fields $p$ and $\rho$ can be expressed as $p = K-V$ and $\rho
= K+V$, where $K= \frac12 \, G_{ij}(\phi) \dot\phi^i \dot\phi^j$
and $V(\phi)$ are the scalars' kinetic and potential energies.
Notice that so long as both $K$ and $V$ are nonnegative, then
$\rho \ge 0$ and $-\rho \le p \le \rho$. Clearly, the condition $p
< - \rho/3$ in this case requires the scalars' evolution at
present to be slow, in the sense that $K < V/2$. The simplest case
in this class is the vacuum energy, for which the scalar field
(and so also $p$ and $\rho$) is time-independent with $p = -\rho$.
In this case the existence of the scalar fields is irrelevant, and
the Dark Energy density is purely attributed to the existence of a
constant vacuum energy (or cosmological constant): $\rho = -p =
V$.

So far so good: although the Dark Energy cannot be a gas of
relativistic or non-relativistic particles, there are
well-motivated kinds of physics whose stress energy can produce a
universal acceleration. Furthermore, it is not too difficult to
choose the functions $V(\phi)$ and $G_{ij}(\phi)$ in such a way as
to account for the known observational features of Dark Energy.
Unfortunately, it has nevertheless proven very difficult to embed
these models into realistic theories of microscopic physics, for
reasons now explained.

\subsection{Modelling Dark Energy}

The next few sections describe some of the issues which make it
difficult to realistically model the universal Dark Energy. The
discussion here loosely follows that of refs.~\cite{DEhard}.

\subsubsection{Scales}

What makes Dark Energy more difficult to understand theoretically
than is Dark Matter is the {\it size} of $K$ and $V$ which are
required, together with some generic features of quantized scalar
fields. This is because any viable scalar-field model must have
two very remarkable properties, which turn out to be quite
difficult to arrange. As is explained below, they must have:
\begin{itemize}
\item {\it Extremely Small Energy Density}, inasmuch as the
present value of the scalar potential must equal the observed Dark
Energy density: $V \sim \rho \sim (3 \times 10^{-3}$ eV$)^4$, and;
\vspace{3 mm}
\item {\it Extremely Small Scalar Masses}, inasmuch as the
scalar's mass must be at most $m_Q \sim 10^{-32}$ eV. This second
condition does not apply if the Dark Energy is simply a
cosmological constant ($K=0$).
\end{itemize}

These requirement are very generic properties of explanations of
the dark energy in terms of a rolling scalar field. Their
necessity may be seen from the scalar field equation of motion
within a cosmological context, which may be written:
\beq \label{KGEq}
    \ddot{\phi}^i + \Gamma^i_{jk}(\phi) \, \dot\phi^j \dot\phi^k
    + 3 \, H \, \dot\phi^i + G^{ij}(\phi) {\partial V
    \over \partial \phi^j} = 0, \qquad \hbox{and}
    \qquad H^2 = {1
    \over 3 M_p^2}\, \left[\frac12 \, G_{ij}(\phi) \, \dot\phi^i
    \dot\phi^j + V(\phi) + \hat\rho \right],
\eeq
where $M_p = (8\pi G)^{-1/2} = 10^{18}$ GeV is the rationalized
Planck mass, $G^{ij}$ is the inverse matrix to $G_{ij}$,
$\Gamma^i_{jk}(\phi)$ is the Christoffel connection constructed by
considering the functions $G_{ij}(\phi)$ to be a metric, and
$\hat\rho$ is the total energy density due to other degrees of
freedom besides $\phi^i$. (Recall, however, that it is $V(\phi)$
which is at present dominating in the universal energy density).
Physically, the quantity $H = \dot a/a$ represents the universal
rate of expansion.

It is instructive to quantify what is required of the potential in
order to obtain a successful description of the Dark Energy. For
this purpose we also restrict our attention to the case of a
single scalar field. Since the scalar potential energy must
dominate during its motion, it is often a good approximation to
suppose that $\phi$ does not move very far from a fixed value,
$\phi_0$, as it moves.\footnote{This simplifying assumption is not
always satisfied, but is also not crucial for the conclusions
being drawn here.} Using the freedom to redefine fields to choose
$\phi_0 = 0$ and $G_{ij}(\phi_0) = \delta_{ij}$ allows us to
expand the functions $G_{ij}$ and $V$ as follows:
\beq \label{Vform1}
    G_{ij}(\phi) = \delta_{ij} + \cdots \,, \qquad\qquad
    V(\phi) = V_0 + \frac{m_\phi^2}{2}  \, \phi^2 + \frac{\lambda}{4}
    \, \phi^4 + \cdots \,.
\eeq
With these choices, we must impose the conditions that the scalar
motion be slow enough to be potential-dominated: $M_p V' \ll V$
and $M_p^2 V'' \ll V$, and that the value of this potential agree
with the observed Dark Energy density: $V_0/M_p^2 \sim H_0^2$,
where $H_0 \sim 10^{-32}$ eV is the present-day universal
expansion rate. As is easily verified, these conditions imply the
two constraints quoted above: $V_0 \sim H_0^2 M_p^2 \sim (10^{-3}$
eV$)^4$, and $m_\phi \ll H_0 \sim 10^{-32}$ eV.

Provided one is happy to choose these values by hand it is
possible to construct phenomenological models of the Dark Energy,
and most of the extant models have this spirit. There are two
separate difficulties with this point of view, one of which is
cosmological in nature and the other of which is to do with
understanding how these numbers arise when we embed the Dark
Energy into a realistic theory of fundamental physics. We briefly
pause to describe the first problem here, before devoting most of
the remainder of this review to the second topic.

\subsubsection{Sensitivity to Initial Conditions}

Although the Hot Big Bang model of cosmology is remarkably
successful in describing what is seen in observations in terms of
a very simple physical picture. It does so, however, only given
very particular initial conditions. For instance it must be
assumed that the initial universe is extremely homogeneous and
spatially very flat. Because evolution within the Hot Big Bang
always makes the subsequent universe less homogeneous and less
flat than it started, it must have been {\it extremely}
homogeneous and flat in the past in order to be as homogeneous and
flat as it presently is (on large distance scales) at such a ripe
old age.

This introduces a worrying sensitivity to a particular set of
initial conditions for the success of the Hot Big Bang
description. One of the main motivations for considering the
possibility that the universe underwent an earlier dramatic
inflationary expansion \cite{inflation} is that it would account
for this initial condition for the later universe in a way which
can be less sensitive to the details of the initial conditions
before this inflationary epoch. This independence of initial
conditions can arise because inflationary configurations are often
{\it attractor} solutions to the system's equations of motion,
inasmuch as a broad class of initial conditions all evolve towards
the attractor solution.

\begin{figure}
  \includegraphics[height=.3\textheight]{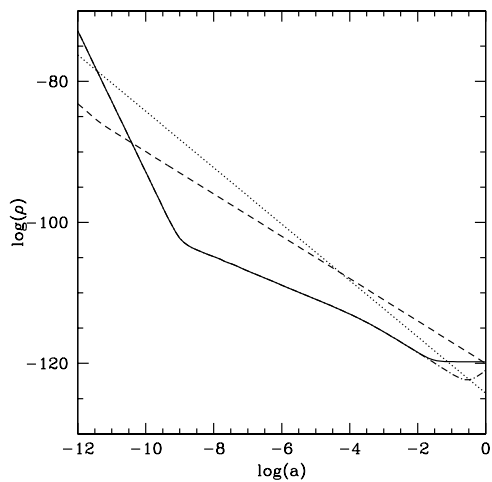}
  \caption{The energy density of radiation (dotted), Dark Matter
  (dashed) and Dark Energy (solid) as a function of universal scale
  factor, $a$, from the nucleosynthesis epoch until now with the
  convention that the present epoch is $a=1$.}
\label{FigOmega}
\end{figure}

A drawback of having a time-dependent Dark Energy is that it seems
to introduce even more sensitivity to initial conditions, and it
is not clear that this new sensitivity can be helped by having an
earlier phase of inflationary expansion. To see why this is so
consider Figure \ref{FigOmega}, which gives the evolution from the
nucleosynthesis (BBN) epoch until the present of the energy
density in radiation, matter and a Dark Energy scalar in the Dark
Energy theory of ref.~\cite{ABRS2}, which we can take to be
representative for these purposes.

As is clear from this figure, the present epoch is very special in
that the energy densities in matter, baryons and Dark Energy are
not so different from one another, even though they were extremely
different in the cosmological past. (Why this should be true is
sometimes called the `Why Now?' problem.) In the model of
ref.~\cite{ABRS2} the coincidence of these energy densities is
ultimately arranged by artfully choosing initial conditions. In
particular the initial value of the Dark Energy field is chosen to
be close to its present value because it tends not to evolve very
far during the epochs after BBN.

One might hope to do better than this, by having the present-day
properties of the Dark Energy emerge in an
initial-condition-independent way, perhaps as an attractor
solution to the relevant field equations. `Quintessence' models
were invented for these purposes, in the hope of explaining why an
evolving scalar field could naturally have an energy density which
is now so similar to that of other forms of matter
\cite{Quintessence}. This hope is based on their equations of
motion admitting `tracking' solutions, within which the scalar
energy density closely follows (or tracks) the dominant energy
density of the universe as it evolves, making it easier also to
understand why the Dark Energy density is presently so close to
the Dark Matter and (to a somewhat lesser extent) radiation
densities. Since these tracking solutions are often also
attractors, they also often describe the late-time evolution of
the scalar field regardless of its initial conditions. The
existence of tracking, attractor solutions therefore opens the
possibility of explaining the present day dark energy in a way
which does not rely on the assumption of special initial
conditions for the quintessence field in the remote past.

Unfortunately, the devil is in the details and since the other
forms of matter do not currently satisfy the Dark Energy equation
of state, $p \lsim -\rho/3$, neither can the dark energy if it is
still in a tracking solution today. So to become the Dark Energy
the scalar must eventually leave the tracking solution. Although
one might hope that this could also be naturally achieved --
perhaps being due to transient behaviour arising from the
crossover from radiation to matter domination
\cite{radmatcrossover} -- so far it has proven difficult to make a
completely convincing cosmology along these lines.

It remains an open problem to see whether or not a time-dependent
Dark Energy can be described in a way which is insensitive to
initial conditions in this way.

\subsection{Naturalness Issues}

The second --- arguably more serious --- problem with Dark Energy
models only becomes evident when one tries to embed the model into
a realistic theory of more fundamental physics. Then one
encounters two different kinds of `fine-tuning' or `naturalness'
problems, which are described in this section and are generic when
scalar fields are used in a microscopic context.

\subsubsection{The Problem of Quantum Corrections}

In order to see why the small values for $V_0$ and $m_\phi$
discussed above are problematic when viewed from a more
microscopic perspective it is worth comparing the potential of
eq.~\eqref{Vform1} with what is normally found for scalar
potentials in microscopic theories.

The theory into which any theory of Dark Energy must be embedded
must also contain the Standard Model of the electro-weak and
strong interactions (or its close approximation). For more than 20
years, this theory has accounted for (and often predicted) the
results of all non-gravitational experiments that have been
performed. The only modification which it has required in this
time is the addition of the neutrino masses which are required to
describe recent discovery of neutrino oscillations. The typical
scale in this theory which sets the masses of most of the
particles appearing in it is the electro-weak scale, $M \sim 250$
GeV.

\begin{figure}
  \includegraphics[height=.1\textheight]{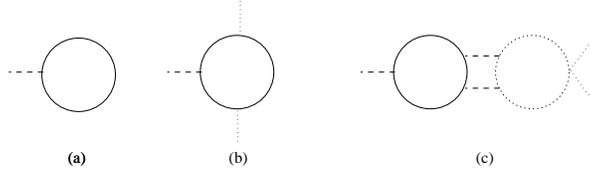}
  \caption{Loop graphs which can generate large quantum contributions
  to the scalar potential.
  The dashed line represents a graviton, a dotted line represents the
  Dark Energy field and a solid line represents a particle of
  mass $M$.} \label{FigLoop}
\end{figure}

The difficulty with a potential like eq.~\eqref{Vform1} arises as
soon as the Dark Energy field, $\phi$, couples to any Standard
Model particles. And these couplings must exist, at least at
gravitational strength because Einstein's Equivalence Principle
requires all fields to couple to gravity. The problem presented by
such couplings is seen once one examines the quantum corrections
to $V(\phi)$ to which they give rise. For instance, since the
strength of couplings to gravity of a particle of energy $E$ are
of order $E/M_p$, loop graphs like those of Figures \ref{FigLoop}
generate the following contributions to $V(\phi)$
\beq \label{DeltaVeq}
    \Delta V_a(\phi) =  \frac{c_0 \, M^4}{(4 \pi)^2}
    \,,\qquad\qquad
    \Delta V_b(\phi) = \frac{c_2 \, g^2 \, M^4}{(4 \pi)^2
    \, M_p^2} \;\; \phi^2 \,,\qquad\qquad
    \Delta V_c(\phi) = \frac{c_2' \, \lambda \, M^6 \,
    }{(4 \pi)^6 \, M_p^4} \;\; \phi^2 \,,
\eeq
where $M$ here denotes the mass of the particle which circulated
within the loop. Figure \ref{FigLoop}b assumes $\phi$ directly
couples to ordinary matter with gravitational coupling strength,
$g \, E/M_p$, while Figure \ref{FigLoop}c assumes only that $\phi$
couples to other fields through the virtual exchange of gravitons.

The factors appearing in these estimates are obtained as follows:
There is a standard factor of $(1/4\pi)^2$ for each loop (in 4
dimensions); and there is a factor of $1/M_p$ for each graviton
vertex (except for the overall tadpole vertex); the factor of
$\lambda$ weights the 4-point $\phi$ vertex coming from the
classical potential; and the power of $M$ follows on dimensional
grounds (in a regularization scheme like dimensional
regularization). Provided that $M$ is the largest mass scale in
the loop, the dimensionless constants, $c_n$, found by explicitly
valuating the graph, are generically ${\cal O}(1)$.

We see from eq.~\eqref{DeltaVeq} that even if the coupling between
the $\phi$ field and the rest of physics is only gravitational in
strength, this coupling generically produces corrections to the
quantities $V_0$ and $m_\phi$ which are bigger than the
phenomenologically-chosen values. Since the corrections grow with
$M$, they are most sensitive to the most massive virtual particles
which can circulate within the loop. For the
phenomenologically-acceptable choice $V_0 \sim (10^{-3}$ eV$)^4$,
Figure \ref{FigLoop}a implies the condition $\delta V_0 \lsim V_0$
is only possible if $M \lsim [(4 \pi)^2 \, V_0]^{1/4} \sim
10^{-3}$ eV. Using Figure \ref{FigLoop}b the condition $\delta
m_\phi \lsim m_\phi$ similarly requires $M \lsim [4 \pi \, m_\phi
\, M_p/g]^{1/2} \sim 10^{-3}$ eV$/g^{1/2}$. Interestingly, for
$g^{1/2} \sim {\cal O}(1)$ the two scales obtained in this way
agree with one another, but they are also clearly many orders of
magnitude smaller than the masses of the known elementary
particles (apart possibly for neutrinos). Using instead Figure
\ref{FigLoop}c and requiring $\delta m_\phi \lsim m_\phi$ gives
the weaker condition $M \lsim 4 \pi \, [m_\phi \,
M_p^2/\lambda^{1/2}]^{1/3} \sim 0.1$ GeV$/\lambda^{1/6}$. Again
(for $\lambda^{1/6} = {\cal O}(1)$) this scale is several orders
of magnitude smaller than the masses of most of the known
elementary particles.

\subsubsection{Two Categories of Fine Tuning}

We see that the quantum corrections to both $m_\phi$ and $V_0$ are
typically much larger than the phenomenologically acceptable
values. On the other hand, only the sum of the classical
contribution and its quantum contributions is ever measured, since
(for example)
\beq
    V_0 = V_0^{cl} + \Delta V_0 \,.
\eeq
Need one really demand that both $V_0^{cl}$ and $\Delta V_0$ be
separately small? Or might they instead both be large, with one
cancelling the other to produce the small observed value?

The existence of such a cancellation is the point of view adopted
in some recent discussions of hierarchies \cite{landscape},
because of the discovery of large numbers of
supersymmetry-breaking vacua within string theory \cite{gkp,kklt}.
The proponents of this attitude notice that different vacua are
likely to produce differently-sized local patches of the universe,
each having a different local value for quantities like $V_0^{cl}$
and $\Delta V_0$. If there are sufficiently many such vacua, and
if the relative probability of obtaining particular values for
$V_0^{cl}$ and $\Delta V_0$ is distributed fairly smoothly, then
there will plausibly be a few vacua for which the cancellation
required to obtain a small $V_0$ occurs. Then our appearance
within such a vacuum is imagined to be explained on {\it
anthropic} grounds, which argue that things like galaxies can only
arise in those regions for which $V_0$ is as small as it is, and
hence it is only in these vacua that living creatures like
ourselves exist \cite{anthropic}.

Although this is a logically defensible point of view, it is
clearly the explanation of last resort. When thinking about the
merits of these arguments I believe it is worth keeping in mind
that hierarchies of scale are also known in many other areas of
physics, and so we may gain insight from these other examples when
thinking about naturalness issues. It is worth rephrasing the
issue within the modern picture of the physics of renormalization,
which underlies our understanding of what the quantum corrections
mean physically \cite{renormalization}.

To this end it is more fruitful not to regard the divide between
$V(\phi)$ and $\Delta V(\phi)$ as a classical/quantum split.
Rather it is more useful to instead think of the `classical'
theory as an effective theory which applies at high energies, and
to think of the quantum corrections as being the contributions
which are obtained as we `integrate out' lower-energy degrees of
freedom to obtain an effective theory which applies at lower
energies. Within such a picture a quantity like $V_0^{cl}$ might
be the result obtained within the effective theory which applies
at energies above the scale $M$ of a particle of interest. The
contribution, $\Delta V$, given by eq.~\eqref{DeltaVeq}, comes
from integrating out this particle to get the effective theory
below the scale $M$ \cite{effectivetheory}. From this perspective
it is hard to see why the contributions of high energies should so
systematically cancel those of low energies, and so the existence
of a small mass scale like $V_0^{1/4}$ or $m_\phi$ raises two
separate questions:
\begin{itemize}
\item {\bf Problem 1:} Why is $V_0^{cl}$ or $m_\phi^{cl}$, so
small at the microscopic scales, say $M \gsim 10^3 \, \hbox{GeV}$,
at which the fundamental theory is couched?
\item{\bf Problem 2:} Why does the quantity $V_0$ or $m_\phi$ {\it
remain} small as all the scales between $M$ and lower energies are
integrated out? (That is, why is $\Delta V_0$ as small as
$V_0^{cl}$, or $\Delta m_\phi$ as small as $m_\phi^{cl}$?)
\end{itemize}
For all of the hierarchies we understand in physics, the smallness
of the hierarchy is understood in both of these ways. That is, we
understand why the corresponding small parameters are small in
{\it both} the high-energy theory, and in the low-energy theory
obtained by integrating out the intervening physics.

For example, there is a hierarchy of about $10^5$ between the
radius of a nucleus and the radius of an atom, and in the
low-energy theory appropriate to atomic physics (QED) this is
understood in terms of the small size of the Bohr wavenumber,
$a_0^{-1} \sim \alpha m_e$, in comparison with the proton mass,
$m_p$. Here $a_0 \gg m_p^{-1}$ because the electromagnetic
coupling, $\alpha = e^2/4\pi$, is small and because the electron's
mass, $m_e$ is small compared with the proton's. If we look to
higher energies, to the theory (QCD) describing physics inside the
nucleus, this hierarchy is understood in terms of the small size
of $\alpha$, and because $m_e$ is much smaller than the QCD scale,
$\Lambda_{QCD}$. Because both $m_e$ and $\alpha$ renormalize
logarithmically, they stay small even as the intervening scales
are integrated out.

{}From this point of view we see that it is {\it conservative} to
also ask that our understanding of the Dark Energy be similarly
understood at all scales. In particular we should ask a reasonable
theory to explain why quantities like $m_\phi$ and $V_0$ are small
both in the effective theory at the TeV scale, {\it and} why they
remain small as all of the known particles having mass $10^{-3}$
eV$ < M < $TeV are integrated out. The recent apparent willingness
to give up this criterion has largely arisen from the desperation
which comes from the great difficulty of doing so in an acceptable
way, given that we know so much about such low-energy scales. The
main message of this section is to recognize that although
anthropic explanations of hierarchies which allow cancellations
between high- and low-energy contributions are defensible, they
are also radical inasmuch as no previous hierarchies we know need
to be understood in this way.

\subsubsection{`Technical' Naturalness}

Of the two naturalness questions posed above, it is the Problem 2
which is the more worrying, because it seems to indicate that we
are missing something in our description of {\it low energy}
physics, which we normally think we understand quite well.
Hierarchies for which Problem 2, above, is understood are said to
be `technically' natural  \cite{technicalnaturalness}. A brief
summary of the known mechanisms for satisfying technical
naturalness for $V_0$ and $m_\phi$ are now summarized.

\paragraph{The Vacuum Energy, $V_0$}

The smallness of $V_0$ has proven to be the hardest to understand
in a technically natural way. Supersymmetric theories provide the
only ray of light in what is otherwise a totally dark picture. On
the one hand, sufficiently many supersymmetries can explain why
the microscopic value, $V_0^{cl}$, must be small or vanish
(thereby addressing problem 1, above). On the other hand, one
supersymmetry can also partially explain why the process of
integrating out lighter particles does not ruin this prediction
(problem 2, above), because supersymmetry enforces a cancellation
between bosons and fermions in their contributions to $V_0$
\cite{susycancellation}. Unfortunately, this cancellation is only
partial if supersymmetry is broken, leaving a residual nonzero
value. If $m_{sb}$ is a measure of the largest mass splittings
between bosons and fermions within a supermultiplet, then the
residual size of the vacuum energy is typically $\Delta V_0 \sim
m_{sb}^2 M^2$, where $M \gsim 1$ TeV is the largest mass in the
problem. For some theories this leading term (for large $M$)
vanishes, in which case the residual result can be as small as
$\Delta V_0 \sim m_{sb}^4$. Sadly, experiment already implies that
$m_{sb}$ must be at least as large as $m_{sb} \sim 10^{2}$ GeV for
observed particles like electrons, so it is has not been clear how
to use this fact in a realistic model. We return to how these
supersymmetric properties might help produce a technically natural
understanding of why $V_0$ is so small in the second half of this
article, when the proposal of Supersymmetric Large Extra
Dimensions (SLED) is discussed in detail.

\paragraph{Cosmologically Light Scalar Fields, $m_\phi$}

There has been more progress in identifying how a small scalar
mass like $m_\phi \sim 10^{-32}$ eV might be technically natural,
with three broad classes of ideas having emerged.

\medskip\noindent {\it 1. Pseudo-Goldstone Bosons:}
The first class of ideas consider the very broad class of scalar
models for which $\phi$ is a pseudo-Goldstone boson (PGB)
\cite{pgb}. A PGB is a (would-be) Goldstone boson for an
approximate symmetry: {\it i.e.} for a symmetry which is both
spontaneously broken (at scale $f$) and explicitly broken (by
terms in the action whose scale is $\mu \ll f$). Since any such
particle must become massless in the limit $\mu \to 0$ for which
the corresponding symmetry becomes exact, for nonzero $\mu$ its
mass is suppressed by powers of $\mu/f$.

Below the scale $f$ the effective equations of motion for any such
a particle has the form of eq.~\eqref{KGEq}, with a kinetic term
and a scalar potential of the form
\beq \label{Vform}
    G_{ij}(\phi) = g_{ij}(\phi/f) \qquad \hbox{and} \qquad
    V(\phi) = \mu^4 \; U(\phi/f) .
\eeq
The emergence of the approximate symmetry as $\mu \to 0$ can be
seen here as the appearance of the shift symmetry, $\phi \to \phi
+ \hbox{constant}$, in this limit.\footnote{This symmetry also
demands a particular form for the kinetic function, $G_{ij}$,
whose precise form depends on the symmetry group.} The existence
of this symmetry is very important, because it implies that any
quantum corrections to $V(\phi)$ must also be proportional to a
positive power of $\mu^4$, and so must again have the form of
eq.~\eqref{KGEq}. This automatically ensures that Problem 2,
above, is solved since the quantum corrections to $m_\phi$ are
naturally as small as is its initial value, $m_\phi^{cl}$
\cite{naturalquintessence}.

To see how this works in detail, notice that in order of magnitude
the scalar potential and its derivatives are given by $V \sim
\mu^4$, $\partial V/\partial \phi^i \sim {\cal O}(\mu^4/f)$ and
$V_{ij} = \partial^2 V/\partial\phi^i \partial\phi^j \sim {\cal
O}(\mu^4/f^2)$. Expanding $g_{ij}$ and $V$ about $\phi = \phi_0$,
and using the freedom to redefine fields to choose $G_{ij}(\phi_0)
= \delta_{ij}$, implies the scalar mass matrix becomes $m^2_{ij} =
{A_i}^k {A_j}^l V_{kl} \sim \mu^4/f^2$, where ${A_i}^k {A_j}^l
g_{kl} = \delta_{ij}$. We see from this that $m_\phi \sim {\cal
O}(\mu^2/f)$, and a similar argument shows that the quartic scalar
self-coupling is $\lambda \sim {\cal O}(\mu^4/f^4)$. For instance,
using this in the above expression for $\Delta V_c$ and taking $M
\sim f \sim M_p$ (more about these choices below) leads to $\Delta
m_\phi = {\cal O}(\mu^2/16 \pi^2 M_p)$, which is indeed smaller
than $m_\phi^{cl} = {\cal O}(\mu^2/M_p)$. (Notice, however, that
the shift symmetry does {\it not} preclude adding a constant to
$V(\phi)$, and so $\Delta V_0$ in this case need not be suppressed
by powers of $\mu$.)

{}From a phenomenological perspective the size of the scales $\mu$
and $f$ is fixed by a successful description of Dark Energy, by
the following arguments (provided the dimensionless functions
$U(x)$, $g_{ij}(x)$ and their derivatives are assumed at present
to be ${\cal O}(1)$). For instance, the value $\mu \sim 10^{-3}$
eV directly follows because by assumption the total scalar-field
potential presently dominates the universal energy density, $\rho
\approx V$. In this way we see that $\mu$ controls the present
Dark Energy density, $\rho \sim \mu^4 \sim (10^{-3} \;
\hbox{eV})^4$, and so also $H \sim \mu^2/M_p$. If it is true that
$K$ is not too much smaller than $V$ then we also know that
$\dot\phi \sim \sqrt{K} \sim \sqrt{V} \sim \mu^2$. The requirement
$f \sim M_p$ is determined from the scalar field equation,
eq.~\eqref{KGEq}, together with the above slow-roll conditions,
since this implies the $\ddot\phi^i$ and the $\Gamma^i_{jk} \,
\dot\phi^j\dot\phi^k$ terms should be much smaller than the other
two. Except for the case of a pure cosmological constant (for
which only the last term is important), we therefore have $H
\dot\phi^i \sim G^{ij} \partial V/\partial \phi^j$ and so
$\mu^4/M_p \sim \mu^4/f$. From this we learn $f\sim M_p$ and hence
$m_\phi \sim \mu^2/f \sim \mu^2/M_p \sim 10^{-33}$ eV.

The virtue of this kind of model is that it shows how to keep the
quantum corrections to the hierarchy between $\mu$ and $f$ --- and
so also the small size of $m_\phi \sim \mu^2/f$ --- small, and so
provide a technically natural solution to Problem 2 for the scalar
mass $m_\phi$. Unfortunately, these arguments shed no light on why
$\mu$ and $f$ should take these values within the high-energy
effective theory in the first place --- {\it i.e.} Problem 1,
above. Nor do they provide any understanding of how to solve
either Problem 1 or Problem 2 for the vacuum energy, $V_0$.

\paragraph{2. Large Kinetic Terms}

A related way to solve Problem 2 for the scalar mass is to have
$m_\phi$ be small due to the largeness of $G_{ij}$ at $\phi =
\phi_0$ rather than due to the smallness of $V_{ij}$ there
\cite{HiZ}. This kind of mechanism relies on the fact that if
$G_{ij} \sim {\cal O}(\epsilon^{-p}) \ll 1$ for $p > 0$ and $0 <
\epsilon \ll 1$, then the matrices ${A_i}^j$ appearing in the mass
matrix, $m_{ij}^2$, are of order ${A_i}^j \sim {\cal
O}(\epsilon^{p/2})$. This has the advantage that even if the
quantum corrections to $G_{ij}$ are large, they can still be
dominated by the still larger contribution of $G_{ij}^{cl}$, and
so their size can be technically natural.

\paragraph{3. Small Cutoffs}

The third proposal which can make $m_\phi$ small enough for Dark
Energy obtains if the 4-dimensional assumptions which underly the
evaluation of the Feynman graphs of Figures \ref{FigLoop} break
down at sufficiently low energies. This can happen if the largest
scale, $M$, for which the effective theory can be 4-dimensional is
as low as $M \sim 10^{-3}$ eV, such as can only happen in Large
Extra Dimensional models \cite{ABRS2,ABRS1}.

As we've seen, such a low value for $M$ could resolve the
4-dimensional part of the technical naturalness issue for {\it
both} $V_0$ and $m_\phi$. In this approach the most difficult
naturalness issues arise in the higher-dimensional effective
theory above the scale $M$, a discussion of which is postponed to
the more detailed description of Supersymmetric Large Extra
Dimensions which follows below.

\subsubsection{The Problem of `Self Tuning': Weinberg's
No-Go Theorem}

There is another difficulty with obtaining a naturally small Dark
Energy density which is related to, but logically distinct from,
the above-discussed issues of quantum corrections and technical
naturalness. This second problem, first articulated in its general
form by Weinberg \cite{ccreview}, arises within those proposals
which use an underlying scale invariance to try to make the the
vacuum energy naturally small. This difficulty is discussed here
in some detail because the Supersymmetric Large Extra Dimensional
proposal of the next section falls into the class of theories to
which Weinberg's objection applies.

Scale invariance frequently arises within Dark Energy proposals
because scale invariance is one of two known symmetries --
together with supersymmetry -- which can force the vacuum energy
to vanish. However, whereas supersymmetry can only force the
vacuum energy to vanish when it is not spontaneously broken, if
scale transformations are a symmetry of the action then the vacuum
energy vanishes whether or not the scale invariance is
spontaneously broken. It can do so because the vanishing of the
vacuum energy is a consequence of conservation of the scale
current and so relies only on the invariance of the action.
Explicitly, the scale transformation property of the metric,
$g_{\mu\nu} \to \ell^2 \, g_{\mu\nu}$, ensures that the
conservation condition for a scale invariant theory is $g_{\mu\nu}
\delta S_{\rm matter}/\delta g_{\mu\nu} = 0$ once all other fields
are evaluated at the solution to their classical equations of
motion. This implies quite generally that any scale-invariant
scalar potential is automatically minimized at a position in field
space for which the potential vanishes: $\left. \partial
V/\partial \phi^i \right|_{min} = V_{min} = 0$. This condition
ensures in turn that $V$ vanishes at this solution, as required.

\begin{figure}
  \includegraphics[height=.2\textheight]{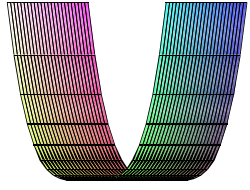}
  \caption{A scale-invariant potential with a flat direction
  along which $V=0$ and scale invariance is broken.}
\label{FigTrough}
\end{figure}

It is useful to see why this works in detail in order to
appreciate Weinberg's objection. For this purpose it is
instructive to consider a collection of scalar fields, $\phi^i$,
which transform under scale transformations in the canonical way:
$\phi^i \to \phi^i/\ell$, for which the sole scale-invariant
potential is quartic:
\beq \label{Vquartic}
    V = \frac{\lambda_{ijkl}}{4} \, \phi^i \phi^j \phi^k \phi^l
    \,.
\eeq
If we choose the $\lambda_{ijkl}$ so that this potential is
strictly non-negative, then it always has a minimum at
$\phi^i_{\rm min} = 0$ for which $V_{\rm min} = 0$. Unfortunately,
the ground state at this point is scale invariant and so the
vanishing of $\phi^i_{\rm min}$ also implies all particle masses
also typically vanish for this vacuum.

A better situation arises if other vacua exist for which
$\phi^i_{\rm min} = v^i \ne 0$, in which case scale invariant
interactions (like ${\cal L}_{\rm yuk} = g_i^{ab}
\overline{\psi}_a \psi_b \phi^i$) can give particles masses which
are proportional to $v^i$. Given any one such a vacuum a
one-parameter family of vacua may always be generated by
performing scale transformations, $\phi^i_{\rm min} = v^i/\ell$.
Since scale invariance is a symmetry (by assumption) we are
guaranteed that all such configurations parameterize a flat
direction of the potential $V$, and since the scale-invariant
point is included as the limiting case $\ell \to \infty$ we also
know that $V = 0$ along this flat direction. This situation is
illustrated in Figure \ref{FigTrough}, and corresponds to the
vector $v^i$ being a zero eigenvector of the matrix
$\lambda_{ijkl}\, v^k v^l$, and shows more explicitly why scale
invariance ensures a vanishing vacuum energy even if it is
spontaneously broken. It is because this offers some hope of
understanding why $V_{\rm min}$ might be much smaller than typical
particle masses, $m \sim g_i v^i$, that scale invariance
frequently arises within proposals for understanding why the Dark
Energy density can be naturally small. Because of the necessity of
having a flat direction, these models also generically predict the
existence of a massless (or very light) scalar particle --- the
dilaton --- which can be regarded as being the Goldstone (or
pseudo-Goldstone) boson for spontaneously-broken scale symmetry.

\begin{figure}
  \includegraphics[height=.2\textheight]{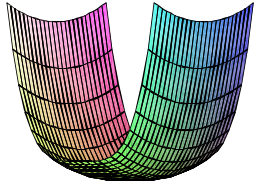}
  \caption{A scale-invariant potential along which the flat direction
  is lifted, with scale invariance not broken at the minimum.}
\label{FigNoTrough}
\end{figure}

We can now also see the weakness of this proposal. Although
spontaneous scale breaking is easily arranged, because it involves
a flat direction we must check that this flat direction is not
lifted by quantum corrections. For internal symmetries the flat
direction which underlies spontaneous symmetry breaking is
normally guaranteed by the symmetry itself but this is {\it not}
so for scale invariance, as can be easily seen for the scalar
example under consideration. Even if we assume there to be no
quantum anomalies to the scale symmetry, the scale invariance of
the quantum action dictates that quantum corrections to $V$ must
have the form of eq.~\eqref{Vquartic}, but with $\lambda_{ijkl}$
replaced by a quantum-corrected coupling, $\lambda^{\rm
eff}_{ijkl}$. But scale invariance by itself does {\it not} impose
additional conditions on these couplings and so in particular
cannot ensure the existence of the flat direction. Generically
even scale-invariant quantum corrections will lift this flat
direction, leading to the situation of Figure \ref{FigNoTrough},
and so although they do not make the vacuum energy nonzero, they
do so by ensuring that the only vacuum is the uninteresting
scale-invariant point: $\phi^i_{\rm min} = 0$.

It is clear that this objection is quite generic, and although it
was encountered in numerous models in the 1980's \cite{80snogo}
Weinberg most effectively emphasized how general it is. At face
value, scale invariance must climb {\it two} hurdles in order to
help with the cosmological constant problem: it must survive
quantization ({\it i.e.} scale anomalies must cancel); and a
separate mechanism must be found which keeps the required flat
direction flat. Otherwise the theory predicts a vanishing vacuum
energy, plus vanishing particle masses of all kinds. Phrased this
way, we can see that supersymmetry may be able to help since (if
it is unbroken) it can be good at producing flat directions.
Unfortunately, the typical curvature of a flat direction which is
generated by quantum corrections in a supersymmetric theory are
also of order the supersymmetry-breaking scale, $m_{sb}$, which
appears to be too large given what is known about elementary
particle masses.

We shall see in the next sections how this last conclusion can be
evaded within special sorts of brane-world models, and indeed the
ability to thread the No-Go result is one of the principal
motivations for this proposal. As is discussed in more detail
below, the main reason these brane-world models can do so is
because in these models the scale of supersymmetry breaking for
the dilaton can be made to be much smaller than the supersymmetry
breaking scale for ordinary particle multiplets. As a result it is
possible for the supersymmetry breaking scale in the dilaton
sector to be as small as the cosmological constant scale,
$10^{-3}$ eV, even though there are no supersymmetric partners for
the observed particles which are light enough to be visible in
particle accelerators.

\section{Supersymmetric Large Extra Dimensions}

The remainder of this article describes a very promising proposal
\cite{Towards,susyADS,Update,MSLED,GGPplus} --- Supersymmetric
Large Extra Dimensions, or SLED for short --- for solving the
cosmological constant problem and understanding the very small
size of the Dark Energy density, $\rho$. This proposal turns out
to predict that the Dark Energy is at present dynamical, and also
naturally addresses the issue of why the rolling scalar fields can
be so incredibly light.

As we shall see, the SLED proposal also has a host of other robust
phenomenological implications besides those it has for cosmology.
These include:
\begin{itemize}
\item Deviations from the inverse square law for gravity, which
more precise estimates show should arise for distances of order
$r/2 \pi \sim 1$ $\mu$m;
\item A particular scalar-tensor theory of gravity at large
distances, with the scalar being the moduli (like the volume)
which describe the two large extra dimensions. This is the same
scalar whose time-dependence now describes the Dark Energy.
\item Distinctive missing-energy signals in collider experiments
at the LHC \cite{susyaddbounds,SLEDscalars} due to the emission of
particles into the extra dimensions.
\item Potential astrophysical signals (and bounds) due to the
possibility of having too much energy loss into the extra
dimensions by stars and supernovae
\cite{PDG,LEDastrobounds,HR,susyaddbounds}.
\end{itemize}
If the SLED proposal is correct, it will be spectacularly so since
it requires this entire suite of observational implications to be
found. Indeed, it is this unprecedented connection between
observables in cosmology and particle physics --- which is driven
by its addressing the fundamental naturalness issues described in
previous sections --- that sets the SLED proposal apart from other
descriptions of Dark Energy.

\subsection{The Proposal}

The two features which define the SLED proposal are large ({\it
i.e.} $r \sim 10$ $\mu$m) extra dimensions, arising within a
supersymmetric theory.

\paragraph{Large Extra Dimensions}

The SLED proposal \cite{Towards,Update}
--- like the LED proposal before it \cite{ADD,ADD2} --- posits that at
present there are two extra dimensions whose circumference, $r$,
is of order $r \sim 10 \, \mu\hbox{m} \sim (10^{-2}$ eV$)^{-1}$.
This is only possible within the framework of the {\it
brane-world} scenario according to which all of the observed
non-gravitational particles are trapped on surfaces (the `branes')
within spacetime, with only the gravitational degrees of freedom
being free to explore the spacetime full extent (the `bulk') away
from these surfaces. The motivation for considering such a
possibility is that it can emerge relatively naturally from some
string theories, such as if the Standard-Model-like spectrum of
massless particles comes from open strings which are localized
near very particular kinds of surfaces (`D-branes') within
spacetime \cite{DBranes}. Although the branes encountered within
string theory tend to have very specific kinds of properties a
more phenomenological approach is adopted here, wherein we try to
identify which brane properties are required to solve the
cosmological constant problem. If any such brane properties can be
identified, then the ultimate goal is to try to see whether branes
having these properties can be found within a more fundamental
theory (like string theory).

A second motivation for working within this kind of brane-world
proposal is the success it has had in recasting low-energy
naturalness issues by identifying unnecessarily restrictive hidden
assumptions which physicists had made about the nature of the
low-energy world. Branes provide the `existence proof' that
reasonable theories of short-distance physics can produce much
more complicated dynamics at low energies than had been hitherto
entertained. The SLED proposal focuses a particular hidden
assumption which might be behind the cosmological constant
problem: the assumption that physics is four dimensional at energy
scales above the cosmological-constant scale, $v \sim 10^{-3}$ eV.
After all, this assumption is crucial to the problematic statement
that a particle of mass $m$ contributes an amount $\Delta V_0 \sim
m^4$. If the world were to involve extra dimensions as large as $r
\sim 1/v \sim 0.1$ mm, then any calculation of the contribution of
a particle having mass $m > v$ (such as the electron) must be done
within a higher-dimensional theory for which the predictions could
differ.

The possibility that extra dimensions can be this large without
being in conflict with experimental observations is based on the
earlier non-supersymmetric large-extra-dimensions (LED) proposal
\cite{ADD}, whose authors first realized that it was possible. The
possibility only arises within a brane-world framework because in
this case only gravitational interactions can probe the existence
of the extra dimensions. In an interesting numerological
coincidence, it happens that the current limit on the size of such
extra dimensions, based on the best present measurements of the
gravitational inverse-square law \cite{eotwash}, is $r/2\pi \lsim
0.1$ mm (which is of order $1/v$).\footnote{Notice that we use the
{\it Jordan frame}, for which $M_p$ is $r$-dependent but the
electroweak scale, $M$, is not. Furthermore, our conventions are
such that $r^n$ is the 2D volume, and so for torii the mass this
implies for a Kaluza-Klein (KK) particle is $M_{\rm KK} \sim 2
\pi/r$.}

In a second extraordinary numerical coincidence, this same radius
also accounts for the small ratio, $M/M_p$, between the weak
scale, $M$, and the Planck scale, $M_p$ --- {\it i.e.} the
electro-weak hierarchy problem. This is because Newton's constant,
$8 \pi G = 1/M_p^2$, is related to $r$ by\footnote{The
supersymmetry breaking scale would be larger -- for fixed $M_p$ --
if the extra dimensions should be warped \cite{susyADS}.} $M_p =
M_g^{1+n/2} \, r^{n/2}$, where $M_g$ is the scale of gravity in
the extra dimensions and $n$ is the number of extra dimensions
whose size is much larger than $1/M_g$. In particular, for $n=2$
this requires the extra dimensions to have $r \sim 1/v$ if the
scale of gravitational physics in the extra dimensions is of order
the electro-weak scale: $M_g = (M_p /r)^{1/2} \sim 10$ TeV $\sim
M$.

\paragraph{Supersymmetry}

SLED also supposes these large extra dimensions to arise within a
supersymmetric field theory, such as might be expected to be
obtained in the low-energy limit of string theory. In such a
picture supersymmetry must be badly broken on our brane, since we
know that there are no super-partners for the observed particles
having masses which are much smaller than $M_g$. Given this scale
for supersymmetry breaking on the brane there is also a
trickle-down of supersymmetry breaking to the `bulk' between the
branes, whose size is set by the bulk's Kaluza-Klein scale,
$m_{sb} \sim M_{\rm KK}$, and so which for unwarped geometries can
be as low as $m_{sb} \sim 1/r \sim 10^{-2}$ eV
\cite{susyaddbounds,Towards}. Much of the success of the SLED
proposal relies crucially on the ability to maintain this
hierarchy between the scales of supersymmetry breaking on the
brane and in the bulk \cite{sepsusyscales}.

\subsubsection{The Action}

In more detail, the action for any SLED variant has the form
\beq
    S_{\rm tot} = S_{SG} + \sum_b S_{b} \,,
\eeq
where $S_{SG}$ describes the action of one of the 6-dimensional
supergravities, and $S_b$ contains the physics on one of possibly
many 3-branes, on of which we find ourselves situated.

\paragraph{The Bulk} The physics of the bulk, $S_{SG}$, is described by
one of the varieties of 6D supergravity. This supergravity can
either be chiral \cite{NS} or non-chiral \cite{Romans}, and can be
gauged or ungauged (in the sense that a particular $R$-symmetry
which does not commute with supersymmetry can arise as a gauge or
a global symmetry). The field content of the 6D gravitational
supermultiplet typically consists of a scalar dilaton ($\phi$),
two symplectic-Weyl spin-1/2 dilatini ($\chi^r, r=1,2$), a 2-form
gauge potential ($B_{MN}$), two symplectic-Weyl gravitini
($\psi_M^r$) and a metric ($g_{MN}$).\footnote{Strictly speaking,
this multiplet is reducible, since a smaller gravity multiplet can
be built using only the metric, gravitino and the self-dual piece
of $B_{MN}$.} Chiral and non-chiral models differ in whether or
not the two bulk dilatini, $\chi$, (or the two bulk gravitini,
$\psi_M$) share the same 6D chirality: $\Gamma_7 \chi^r = \pm
\chi^r$.

Additional matter multiplets can also be present, including
hyper-multiplets (2 real scalars, $\Phi^i$, and a symplectic-Weyl
spin-1/2 particle), gauge multiplets (a gauge boson, $A^a_M$, plus
2 symplectic-Weyl spin-1/2 fermions), tensor multiplets, massive
KK multiplets, and so on. For the present purposes we take the
gauged, chiral 6D theory coupled to gauge multiplets as being
representative, for which the bosonic action is
\beq \label{E:Baction}
     \frac{{\cal L}_{SG}}{e_6} = -\, \frac{1}{2 } R - \frac{1}{2 }
    \partial_{M} \phi \, \partial^M\phi  - \frac12 \,
    G_{ij}(\Phi) \, \partial_M \Phi^i \, \partial^M \Phi^j -
    \frac{e^{-2\phi}}{12} \; G_{MNP}G^{MNP} -
    \frac{e^{-\phi}}{4} \; F^a_{MN}F_a^{MN} - {v(\Phi) e^\phi},
\eeq
in units for which $M_g = 1$: {\it i.e.} $M_g^{-4} = 8 \pi G_6 =
1$. As usual, in this expression $e_k = \sqrt{-\det g_{MN}}$ in
$k$ spacetime dimensions. The hypermultiplet potential, $v(\Phi)$,
has a minimum at $\Phi^i = 0$ for which $v(0) = 2 g^2$, where $g$
denotes the gauge coupling of the specific $R$ symmetry for which
none of the bosons listed above transforms (although their
superpartners do).

\paragraph{The Branes}
The cosmological-constant mechanism described below does not
depend on many of the details of $S_b$. For phenomenological
purposes it should be emphasized that because our brane breaks
supersymmetry, this symmetry is only nonlinearly realized
\cite{nonlinearsusy} and in particular does not imply the
existence of superpartners for ordinary particles in the usual
sense. Rather, a supersymmetry transformation of a light brane
state (like the electron) instead produces a massless Goldstone
fermion, which is eaten by the massless Kaluza-Klein mode of the
bulk gravitino to give this particle its mass. Since
nonlinearly-realized gauge symmetries (like supergravity) are
indistinguishable at low energies from explicitly broken ones
\cite{UsesandAbuses}, the result is an effective theory of
brane-bound states at TeV energies which appears to explicitly
break supersymmetry \cite{nonlinearbranesugra}.

In the minimal SLED version (MSLED) one makes the particularly
simple choice that the action on our brane is given (apart from
terms which couple brane and bulk fields) by precisely the
Standard Model \cite{MSLED}, but although this is particularly
predictive in its implications for collider experiments it is not
a required choice.

For reasons which will become clear in what follows, it will turn
out in many cases that the bulk dilaton, $\phi$, is related to the
extra-dimensional volume by $e^\phi \sim (M_g r)^{-2} \ll 1$, and
so we must also suppose that gauge couplings amongst the physical
fields on our brane do not vanish in the limit that the bulk
dilaton runs away to asymptopia: $e^\phi \to 0$. That is, if the
physical coupling of brane-bound gauge fields is denoted $g_b$,
then we must demand $g_b = g_b^0 + g_b^1 \, e^\phi + \cdots$, with
$g_b^0 \ne 0$. As we shall see below, arranging for this to be
true in the Einstein frame may be difficult to accomplish within
string theory, depending on the relation between the 6D dilaton
and the underlying string coupling.

\subsection{Classical Relaxation of the 4D Cosmological Constant}

Within the above framework gravitational physics is effectively
6-dimensional for any energies above the scale, $1/r \sim 10^{-2}$
eV, and so the cosmological constant problem must be posed within
this new context. In order to see how the cosmological constant
problem is phrased in 6 dimensions, one must integrate out the
degrees of freedom between the scales $M_g \sim 10$ TeV and $1/r
\sim 10^{-2}$ eV. We seek the cosmological constant within the
effective 4D theory obtained after performing this integration,
which describes gravitational physics (like present-day cosmology)
on scales much larger than $r$. Imagine, therefore, performing the
integration over modes having energies $1/r < E < M_g$ in the
following three steps \cite{Towards}:
\begin{enumerate}
\item First, integrate out (exactly) all of the degrees of freedom
on the branes, to obtain the low-energy brane dependence on the
massless 4D graviton mode. In so doing we obtain (among other
things) a large effective brane tension, $T \sim M_g^4$ for each
of the 3-branes which might be present, which includes the vacuum
energies of all of the presently-observed elementary particles.
\item Next, perform the classical part of the integration over the
bulk degrees of freedom. This amounts to solving the classical
supergravity equations to determine how the extra dimensions curve
in response to the brane sources which are scattered throughout
the extra dimensions. It will be argued that it is this classical
response which cancels the contributions from the branes obtained
in Step 1, above.
\item Finally, perform the quantum part of the integration over
the bulk degrees of freedom. Given the cancellation of the
previous two contributions, it is this contribution which is
responsible for the fact that the present-day Dark Energy density
is nonzero. It is argued below that in some circumstances this
quantum contribution is of order $m_{sb}^4$, where $m_{sb} \sim
M_g^2/M_p \sim 10^{-2}$ eV is the supersymmetry-breaking scale in
the bulk. The small size of the 4D vacuum energy is in this way
attributed to the very small size with which supersymmetry breaks
in the bulk relative to the scale with which it breaks on the
branes.
\end{enumerate}

In a 4D world, the only contribution we would have is that of Step
1, above, and the problem is that this is much too large. But in a
6D world, because all of the observed particles are localized on
our brane their vacuum energy should be thought of as a localized
energy source in the extra dimensions, to which the bulk geometry
must respond. The next section argues that the classical part of
this bulk response (Step 2, above) is of the same order as the
direct contribution of Step 1, and precisely cancels it in a way
which does not depend on the details of the supergravity involved
or of the precise extra-dimensional geometry which lies between
the various branes. In this way it provides a 6 dimensional
realization of self-tuning, whereby the effective 4D cosmological
constant is automatically adjusted to zero by the classical
response of the 2D bulk to the brane sources. The size of quantum
bulk contributions are discussed in more detail below, after first
pausing to discuss the nature of this classical self-tuning.

\subsubsection{6D Self-Tuning: A Cartoon}

Step 1 consists of exactly integrating over all brane fields
having masses larger than $1/r$, and this produces a variety of
local interactions in the effective theory for energies $E \lsim
1/r$ on the brane. Since our interest is in the dependence of the
effective theory on the 4D metric, and we assume a large volume
for the extra 2 dimensions -- $M_g r \gg 1$ -- we may expand these
effective interactions in powers of the curvature:
\beq
    \delta {\cal L}_b = - \sqrt{-g} \, \left[ \delta_b T_b +
    \frac12 \, \delta_b \mu_b^2 \, R + \cdots \right]\,,
\eeq
where on dimensional grounds we expect $\delta_b T \sim M_g^4$,
$\delta_b \mu_b^2 \sim M_g^2$ {\it etc}..\footnote{The subscript
`$b$' on $\delta_b$ here distinguishes these from similar, but
independent, contributions to the localized brane action which
arise from integrations over short-wavelength bulk modes ({\it
i.e.} when performing Steps 2 and 3).}

Step 2 consists of the classical part of the bulk integration, and
so is equivalent to substituting into the classical action the
bulk field configurations which are found by solving the classical
field equations using the above effective brane action as a
source. It happens that for many situations this classical
response of the extra dimensions can be computed explicitly within
the approximation that the branes are regarded as delta-function
tension sources. Because 3-branes in 6 dimensions have
co-dimension 2, the discussion resembles the better-known analysis
of the gravitational field of a cosmic string in 4 dimensions
\cite{cosmicstring}.

For co-dimension 2 the extra-dimensional curvature tensor
typically acquires a delta-function singularity at the position of
the branes, corresponding to a singularity at the brane
position.\footnote{The generality of this form of singularity is a
question which is examined more closely in later sections,
following the discussion of ref.~\cite{GGPplus}.} Einstein's
equations require that the singular contribution to the
two-dimensional curvature given by
\beq \label{Ricci2}
    \sqrt{g_2} \, R_2 = - 2\, \sum_b T_b \,
    \delta^2(y-y_b) + \hbox{(smooth contributions)} \, ,
\eeq
Here $y_b$ denotes the position of the `$i$'th brane in the
transverse 2 dimensions, and the `smooth contributions' are all of
those which do not involve a delta-function at the brane
positions.

The effective 4D cosmological constant after performing Steps 1
and 2 above is obtained by plugging the above expression into the
classical bulk action, eq.~\eqref{E:Baction}. The effective 4D
cosmological constant obtained at this order is then
\beq \label{rhocl}
    \rho_{\rm cl} =  \sum_b T_b + \int_M d^2y \; e_2 \,
    \left[\frac12 \, R_2 + \dots \right] \nonumber \\
    = 0 \, ,
\eeq
where the sum on `$b$' is over the various branes in the two extra
dimensions and `$\dots$' denotes all of the other terms besides
the Einstein-Hilbert term in the supersymmetric bulk action. The
final equality here has two parts. First, the sum over brane
tensions, $T_b$, precisely cancels the contribution of the
singular part of the curvature, eq.~\eqref{Ricci2}, to which they
give rise \cite{CLP}. Second, a similar cancellation also occurs
amongst the various `smooth' contributions in $\rho_{\rm cl}$ once
these are evaluated for all of the bulk fields using the classical
field equations \cite{Towards}. Interestingly, this cancellation
does not depend on the details of the bulk geometry, or on the
number of branes, since it relies only on a classical scale
invariance which all 6D supergravity actions enjoy \cite{susyADS}.
Best of all, this cancellation does not depend at all on the {\it
value} of the brane tension, $T_b$, and so applies equally well
even if these tensions are large and include all of the quantum
effects due to virtual particles localized on the branes.

We are left with the contribution of quantum effects in the bulk
(Step 3), to which we return in more detail in subsequent
sections. These must ruin the brane-bulk cancellation because the
scale invariance of the classical supergravity equations is not a
bona-fide quantum symmetry. However the bulk sector of the theory
is also one which is almost supersymmetric, since the bulk
supersymmetry-breaking scale is very small: $m_{sb} \sim 1/r \sim
10^{-2}$ eV. As a result we might expect standard supersymmetric
cancellations to suppress the quantum part of the result by powers
of $m_{sb}^2$, and if the leading term should be of order
$m_{sb}^4$ this would be the right size to account for the
observed Dark Energy density. As is shown in later sections, under
certain circumstances this is indeed what happens for some 6D
supergravities.

Before evaluating the quantum contributions of Step 3, we first
take a more detailed look at the classical 6D self-tuning
mechanism which was just described.

\subsubsection{6D Self-Tuning I: General Solutions}

The very name `self-tuning' carries a whiff of swindle, partly
because previous self-tuning proposals \cite{5dst} were later
found really to be cosmetic fixes \cite{5dnotst}, which merely hid
the tuning of the cosmological constant to zero. It is natural to
suspect a similar swindle is occurring here, and so the present
section describes some of the recent steps which have been taken
to rule out hidden fine-tunings within the classical part (Steps 1
and 2) of the argument given above. These steps were made possible
by the explicit construction of a very general class of solutions
to the chiral, gauged supergravity in 6 dimensions \cite{GGP}.
These solutions are now summarized, with a view to seeing whether
or parameters need be specially adjusted in order to keep the
effective 4D cosmological constant zero.

\paragraph{Explicit Solutions}

The simplest explicit solutions to 6D chiral, gauged supergravity
in the presence of source branes is the `rugby-ball' solution of
ref.~\cite{Towards},\footnote{Similar solutions within
non-supersymmetric theories are considered in
ref.~\cite{RB-nonsusy}.} which consists of product-space metric
\beq  \label{rugbymetric}
    \exd s^2 =  \eta_{\mu\nu}(x) \, \exd x^\mu \exd
    x^\nu + \exd s^2_2 \,,
\eeq
where $\eta_{\mu\nu}$ is the Minkowski-space metric and $\exd
s^2_2 = r^2(\exd \theta^2 + \sin^2\theta \, \exd \varphi^2)$ is
the line element on a 2-sphere of radius $a$. This solves the
supergravity equations obtained from eq.~\eqref{E:Baction}, with
constant dilaton, $\phi_0$ and a Maxwell field strength, $F_{mn} =
f \epsilon_{mn}$, which is proportional to the 2D volume form. For
these choices the field equations imply the parameters $\phi_0$,
$f$ and $r$ must satisfy $g^2 e^{\phi_0} \propto gf \propto
1/r^2$. The parameter $r$ is undetermined by the field equations,
and parameterizes the flat direction which is guaranteed to exist
at the classical level due to the classical scaling symmetry of
the supergravity field equations.

Branes are incorporated into this solution by removing a wedge
from the 2-sphere whose edges are lines of longitude and
identifying these edges once the wedge is removed. This removal
and identification introduces a conical singularity to the North
and South poles of the 2-sphere, which represent the gravitational
back-reaction of the branes onto the bulk geometry. Notice in
particular that because the defect angles at the north and south
poles are the same size for the rugby-ball solution, the tension
of each of the branes situated at these points must be equal to
one another. We wish to ascertain whether or not these kinds of
constraints constitute the hidden fine-tunings which are
responsible for keeping the 4 dimensions flat. More general
solutions to the 6D supergravity field equations --- for both the
chiral and non-chiral theories --- have also been found
\cite{susyADS}, as well as new kinds of solutions to the
non-supersymmetric theories \cite{Seif}, for which the large 4
dimensions are warped and the small internal 2 dimensions are
squashed.

The best test of the above self-tuning argument at present
available is based on a class of solutions to 6D chiral, gauged
supergravity obtained in ref.~\cite{GGP} by Gibbons, Guvens and
Pope (henceforth GGP). What makes these solutions so useful as a
test of self-tuning is that these authors derive the {\it most
general} solution to these field equations subject to two
assumptions: ($i$) maximal symmetry in 4 dimensions ({\it i.e.} de
Sitter, Minkowski or anti-de Sitter space), and ($ii$) axial
symmetry in the internal 2 dimensions. That is, they find the most
general solutions whose metric has the form
\beq  \label{GGPmetric}
    \exd s^2 =  W^2(\theta) g_{\mu\nu}(x) \, \exd x^\mu \exd
    x^\nu + \exd\theta^2 + a^2(\theta) \exd \varphi^2 \,,
\eeq
and for which $\phi = \phi(\theta)$ and $A_\phi = A_\phi(\theta)$.
Here the intrinsic 4D metric, $g_{\mu\nu}$, satisfies
$R_{\mu\nu\lambda\rho} = c (g_{\mu\lambda} g_{\nu\rho} -
g_{\mu\rho} g_{\nu\lambda})$, for some constant $c$. The assumed
axial symmetry corresponds to shifts of the coordinate $\varphi$,
and the metric can have singularities at up to two positions,
$\theta = \theta_\pm$, within the internal 2 dimensions
corresponding to the positions of source branes. GGP find that
there is a five-parameter family of solutions to the supergravity
equations subject to these symmetry conditions. What is most
remarkable about these solutions is that {\it every single one of
them has a flat intrinsic 4D geometry} ({\it i.e.} $c = 0$), even
though none of them is supersymmetric (except for the single
spherical solution with no wedge removed, mentioned above), as is
consistent with the expectation that the effective 4D cosmological
constant vanishes.

Given a class of general solutions it becomes possible to begin to
address the question of whether hidden fine-tunings are
responsible for this 4D flatness. In particular, are the brane
tensions and couplings being carefully adjusted by an unseen
sleight of hand? A partial answer to this question is now
possible, and although at first sight such an adjustment appears
to take place, it is now explained why these appearances may be
deceiving \cite{GGPplus}.

In order to do so it is useful to keep track of the physical
meaning of the parameters on which the general GGP solutions
depend. There are 5 such parameters, but one of these simply
parameterizes the flat direction whose existence is guaranteed by
the classical scale invariance of the supergravity equations. A
second parameter corresponds to another classical scaling
property, under which a redefinition of the fields may be used to
rescale the gauge coupling $g$ to any positive fixed value. The
three remaining parameters are broadly related to the three
physical quantities which characterize these geometries: the
tensions, $T_\pm$, of the two branes which source the bulk
geometry; and the overall magnetic flux of the background magnetic
field which (marginally) stabilize it.

We now ask what may be said about the natural of self-tuning given
the properties of these general solutions. What is at first sight
disturbing is that for the generic GGP solutions these parameters
are {\it always} subject to a topological constraint, and those
solutions for which the metric singularities are purely conical
are subject in addition to a second, nontopological constraint
\cite{susyADS,GGPplus}. Since either (or both) of these
superficially might be thought to be the hidden self-tuning, we
next discuss each of them in turn.

\paragraph{Topological Constraints}

The only constraints which hold quite generally for the GGP
solutions are those which have their origins in topology. In fact,
there are two topological conditions which these solutions all
share: one which expresses that the internal 2D geometry is
topologically a sphere; and one which expresses the quantization
(and conservation) of magnetic monopole flux
\cite{Towards,GibbonsPope}.\footnote{See also \cite{RB-nonsusy}
for a similar discussion in the non-supersymmetric case.} Since
the first of these turns out to hold for all values of the
parameters describing the classical solution \cite{GGP}, it is of
less interest as a potential source of fine-tuning. The
quantization of monopole number potentially has more to say in
this regard, since it directly imposes a relation between the
brane tensions, the gauge couplings and one of the 5 parameters
characterizing the background solutions.

The resulting topological constraint can be written in the
following way \cite{GGPplus}:
\beq \label{tensiontopology}
    \frac{g^2 e^{-\phi_0/2}}{2} \left( \frac{T_+ - T_-}{4 \pi}
    \right) = N^2 \left( \frac{g^2}{\tilde{g}^2} \right) \,,
\eeq
where $N$ is the integer which labels the monopole number. Here
$T_\pm$ are the two brane tensions, $g$ is the gauge coupling
which appears explicitly in the 6D supergravity action,
eq.~\eqref{E:Baction}, $\tilde g$ is the gauge coupling for the
background magnetic field and $\phi_0$ is an additive constant in
the dilaton configuration (and so is one of the parameters
describing the solution). The limit of equal tensions, $T_+ \to
T_-$, of this expression involves a subtlety, because one must
also take $g^2 e^{-\phi_0/2} \to \infty$ --- with the product
$(T_+ - T_-) \, g^2 e^{-\phi_0/2}$ fixed --- in order to keep the
explicit solution well-defined. If both limits are taken then the
rugby-ball solution discussed above is obtained, with $T_\pm =
4\pi(1 - k)$ for some $k$. In this case the above topological
constraint degenerates to $k^2 = N^2 (g^2/\tilde{g}^2)$.

It is tempting to conclude that this monopole-number constraint is
the hidden fine-tuning we seek, since it explicitly requires the
brane tensions appearing in the solution to be adjusted relative
to one another. But first impressions deceive, and this initial
conclusion does not stand up to closer scrutiny, as is now argued
(following refs.~\cite{susyADS,Update}). Recall in this regard
that the crucial issue for fine-tuning is whether or not the
constraint is stable against renormalization. That is, if
eq.~\eqref{tensiontopology} is imposed amongst the renormalized
quantities at the TeV scale, does it automatically remain imposed
as successive scales are integrated out down to the scales below 1
eV? If so, then the constraint is technically natural, in the
sense described above, and so is not fine-tuned this (the most
serious) notion of fine tuning. But topological constraints are
{\it always} natural in this sense, because the integrating out of
successive scales of physics is a continuous process, and
precisely because topological constraints involve quantization of
quantities in terms of integers, they remain unchanged by any such
continuous process. Although the functional form of the
topological constraint in terms of the renormalized parameters of
a theory may change with scale, the validity of the topological
constraint itself is guaranteed, by continuity, to be stable under
renormalization.

So what do these topological constraints mean? Their meaning is
probably most clear if the branes are allowed to couple directly
to the background electromagnetic flux, through a term in the
brane action of the form \cite{susyADS}
\beq
    \Delta S_b = q_b \int {}^\star F \,,
\eeq
where ${}^\star$ denotes the 6D Hodge dual of the 2-form field
strength. (Such a coupling is not turned on for the GGP solutions
under discussion.) In the presence of such couplings the
topological constraint of eq.~\eqref{tensiontopology} instead
states that the difference $q_+ - q_-$ is proportional to $
\frac12 \, {g^2 e^{-\phi_0/2}} \left( {T_+ - T_-} \right) /{4 \pi}
- N^2 \left( {g^2}/{\tilde{g}^2} \right)$, which reduces to
eq.~\eqref{tensiontopology} in the limit where $q_+ = q_-$. Seen
in this way, the topological constraint restricts the kinds of
magnetic couplings which are possible on topological grounds, in
much the same way that topology only allows electrostatic
potentials to be constructed for a collections of charges within a
compact space, if the sum of all the charges vanishes. That is, it
is the integrability condition for the existence of a solution to
the field equations subject to the given boundary conditions.

\paragraph{Conical Singularities}

Although only the topological constraints holds quite generally
for all of the GGP solutions, there is another issue which may
bear on whether or not the solutions involve a hidden fine tuning
\cite{GGPplus}. This additional issue concerns the kinds of metric
singularities to which physically reasonable 3-branes can give
rise within 6D supergravity. In order to understand the issue,
recall the discussion of the gravitational field of a cosmic
string in 4 dimensions (using pure Einstein gravity)
\cite{cosmicstring}. In this situation there are two solutions to
the Einstein equations which have the symmetries required to be
the field of a cosmic string. One of these is flat space, and the
other is a curved space whose geometry resembles the shape of the
mouth of a trumpet \cite{newcosmicstring}. Of these, the flat
solution exhibits a conical singularity at the string position (in
the limit of an infinitely thin string), while the trumpet-shaped
solution has a real curvature singularity there. In principle,
which solution applies in any physical situation can be determined
by resolving the internal structure of the string, in order to
smooth out the metric at the string's center. This can be done
explicitly, for example, in terms of gauge and scalar fields if
the string is a weakly-gravitating Nielsen-Oleson string within a
gauge theory. In this case it can be proven that only the flat
solution (with conical singularity) can be smoothly joined to the
smooth geometry of the string's interior \cite{Ruth}.

{\it If} it is true that only conical singularities can also be
produced by physically reasonable 3-branes in 6 dimensions, then a
hidden fine-tuning might underly the vanishing 4D cosmological
constant within 6D supergravity. This is because only a subset of
the general GGP solutions involve purely conical
singularities,\footnote{For all of the GGP solutions the 2D metric
in polar coordinates centered at a brane can be written in the
form $\exd r^2 + c \, r^p \, \exd \varphi^2$ as $r \to 0$, for
constants $c$ and $p$. The singularity at the brane position is
only conical if $p = 2$, in which case $c$ is related to the
defect angle at the cone's apex \cite{GGPplus}.} defined by the
family of GGP solutions whose tensions satisfy the condition
\cite{susyADS,GGPplus}:
\beq \label{Tconstraint}
     \left(1 -
    \frac{T_+}{4\pi} \right) \left( 1 - \frac{T_-}{4\pi} \right)
    = \frac{g^2\, e^{-\phi_0/2}}{2} \left( \frac{T_+ - T_-}{4 \pi}
    \right) = N^2 \left( \frac{g^2}{\tilde{g}^2} \right) \,.
\eeq
Here the last equality follows from using the topological
constraint, eq.~\eqref{tensiontopology}. (Recall that the limit of
equal tensions -- the rugby ball -- requires taking the limit $g^2
\, e^{-\phi_0/2} \to \infty$ in such a way that the middle
expression of eq.~\eqref{Tconstraint} need not vanish.) We see
that conical singularities are only possible for a one-parameter
locus of tensions within the $T_+ - T_-$ plane. A key question
then becomes whether or not there is a physical reason why
solutions must be chosen having only conical singularities.

Must GGP solutions satisfy eq.~\eqref{Tconstraint} in order to
make physical sense? Possibly not, although a definitive
resolution of this question is not yet possible. We must ask: Can
solutions having non-conical singularities be produced by any
reasonable resolution of the 3-brane structure? There are two
reasons for entertaining the possibility of solutions having more
complicated singularities than conical in the 6D context.
\begin{itemize}
\item For supergravity the existence of other fields in the bulk
--- such as the dilaton and other superpartners of the metric ---
can complicate the singularity structure since these other fields
can themselves diverge as the brane position is approached
\cite{gregory}.
\item For cosmic strings in 4D the assumption of weak
gravitational fields plays an important role in concluding that
the external geometry is flat, and this need not be true for
3-branes in 6 dimensions.
\end{itemize}
The establishment of whether or not all of the general GGP
solutions can be sourced by physically reasonable branes is
clearly an important issue to resolve for deciding whether or not
a hidden tuning has been slipped into the SLED proposal, and is
currently under active study.

\subsubsection{6D Self-Tuning II: Response to Tension Changes}

There is another way for which the properties of the GGP solutions
can help understand whether or not 6D supergravity really provides
a dynamical mechanisms for relaxing the effective 4D cosmological
constant. In particular -- as was first discussed in
ref.~\cite{Update}, and again in more detail in
\cite{GGPplus}\footnote{See also ref.~\cite{GP}.} -- if successful
the SLED proposal must ultimately explain why the small effective
4D cosmological constant seen by brane observers is robust against
arbitrary changes to the various brane tensions, such as might
occur due to phase transitions taking place on the branes.

Imagine, then, we start off with one of the GGP geometries,
perhaps one having only conical singularities, for which brane
observers experience a flat 4 dimensions. Suppose that at time
$t=0$, one of these tensions, say $T_-$, is instantaneously
locally perturbed to a new value, $T'_-$, and then held fixed and
that the other tension, $T_+$, does not change during this
process. How does the bulk geometry respond?

An honest calculation of this response requires a fully
time-dependent treatment, because causality guarantees that the
influence of the tension change can only propagate into the bulk
at the speed of light. Our interest is in the final state towards
which this transient evolution heads, and it is easiest to think
about these alternatives by thinking of the 6D theory as a
complicated 4D theory consisting of a great many massive
Kaluza-Klein modes, $\Phi_k$, whose vacuum values are determined
by a complicated multi-dimensional scalar potential,
$V_{KK}(\Phi_k)$, which could be computed (in principle) by
dimensional reduction. Our initial GGP solution may be regarded as
being a local minimum of this potential at which $V_{KK} =
0$.\footnote{More properly, a GGP (or any other classical)
solution corresponds to a point along a flat trough of this
potential, due to the existence of the marginal flat direction
which is always guaranteed at the classical level by the classical
scale invariance of the 6D supergravity equations.} Since this
potential depends on the values of the brane tensions it is
perturbed once these tensions are adjusted, and the time-dependent
evolution describes the rolling of the various KK modes towards
their new minima. There are essentially three possibilities which
are possible for the endpoint of this time-dependent evolution:
\begin{enumerate}
\item {\sl Successful Self-Tuning:} In this picture, the endpoint
of the transient evolution is another static geometry, for which
the intrinsic 4D brane geometry is flat. In this case the
evolution ends at a new local minimum of $V_{KK}$ which also
vanishes, such as would occur if the new minimum corresponded to
another GGP solution.
\item {\sl Non-catastrophic Adjustment:} In this picture, the
endpoint of the transient solution is another geometry for which
the internal 2 dimensions are locally stable (modulo the flat
direction), but for which $V_{KK}$ has been perturbed to be
nonzero. In this case we would expect there being another solution
to the 6D field equations corresponding to always being in this
new 2D geometry, but with the 4 dimensions being described by de
Sitter or anti-de Sitter space.
\item {\sl Catastrophic Runaway:} A final option is a runaway,
wherein the time-dependent evolution might shrink the 2D geometry
to a point, or inflate it to infinite size. From the 4D
perspective this would correspond to $V_{KK}$ developing a new
unstable direction, along which the field can roll without
stopping. Notice that this runaway option {\it cannot} arise (at
the classical level) as a destabilization of the original flat
direction, since the flatness of this direction is always
guaranteed by the underlying scale invariance at the classical
level now being discussed.
\end{enumerate}

Although the results of a full time-dependent analysis are not yet
available, some conclusions can be drawn given the existence of
the general GGP solutions. In particular, Option 2 above can be
ruled out for time evolution which does not break the initial
axial symmetry of the extra dimensions. If the endpoint were
towards a locally stable, axially symmetric 2D geometry having
nonzero $V_{KK}$, then we'd expect there to also exist solutions
to the 6D equations corresponding to eternally living at this 2D
geometry, with the intrinsic 4-geometries being either de Sitter
or anti-de Sitter spaces. But the GGP results show that such
solutions do not exist, allowing this option to be ruled out.

It is also clear that since time evolution is continuous, the
topological constraints (monopole number and Euler number
satisfied by the initial configuration are guaranteed to remain
preserved during the transient evolution. Of course, the
particular expressions of these constraints as functions of the
parameters of the initial solutions might change, since the new
solutions might not satisfy the earlier assumptions under which
the consequences of the topology was computed (such as by not
being axially symmetric). But the topological nature of the
constraints ensures that they automatically remain satisfied by
the final solutions, as well as the time-dependent solutions which
lead to them. Similarly, conserved quantities like energy and
angular momentum must also remain unchanged during the
time-dependent evolution.

It is clearly of interest, then, to know if solutions within the
GGP class exist for {\it any} choice of topological flux and
tensions, since if such a solution were to exist then we might
reasonably expect that the endpoint of the time-dependent
evolution would be towards this static solution. This would
correspond to Option 1 above, because the various KK modes would
all settle down into new minima of their potentials at a point
where $V_{KK} = 0$ by virtue of the 4D flatness of all of the GGP
solutions.

Given our present knowledge, we {\it can} say something about
whether or not GGP solutions exist for pairs of tensions $T'_-$
and $T_+$, at least for those which are small perturbations of an
already existing GGP geometry. We can do so because it is possible
to compute the brane tensions and magnetic flux of the source
fields as functions of the 3 parameters of the GGP solutions which
survive once the two scaling relations are used to set an additive
constant in the dilaton, $\phi_0$, and the $U(1)_R$ coupling $g$
to specific values. A calculation of the Jacobian of these
expressions shows that they are invertible, and so a choice of GGP
parameters exists for any pair of tensions in the neighborhood of
a given GGP solution \cite{GGPplus}. In this way we see that it is
plausible that the endpoint of the transient evolution could be
towards another GGP solution, for which the 4D geometry is flat.

Clearly the endpoint of the evolution of the solutions once their
source tensions are changed can only be properly resolved by
constructing explicitly time-dependent solutions to 6D
supergravity. Such a construction is underway, and so a definitive
answer should soon be at hand \cite{timedependent}. In the
meantime one might ask, why work so hard to answer this question
in 6 dimensions at all, when the important part of the analysis
could perhaps be done within a simpler 4D effective treatment?

\subsubsection{6D Self-Tuning III: A 4D Analysis?}

A 4D analysis of this type has been performed \cite{GP}, which
studies self-tuning in 6 dimensions, both for theories with and
without supersymmetry, leading to the claim that the runaway
solution (Option 3 above) is the one chosen by the theory. For the
non-supersymmetric case this claim supports earlier, more
detailed, studies \cite{nonsusycase} who also find difficulties
with self-tuning within the non-supersymmetric context. It also
agrees with the arguments given above inasmuch as the required
underlying classical scale invariance is absent in the various
non-supersymmetric 6D proposals. But does this analysis also rule
out self-tuning within 6D supergravity, where such a scale
invariance certainly exists? In this section it is argued that the
calculations of ref.~\cite{GP} are insufficient to tell.

The key assumption of the quantitative part of their analysis is
the ansatz is to truncate the full 6D equations on the rugby-ball
solutions to 4 dimensions, to restrict attention to two particular
fluctuations about this solution: the time-dependence of the
extra-dimensional volume, $a = e^{\psi(x)}$, and the 6D dilaton,
$\phi = \phi(x)$. The background Maxwell field also is taken
proportional to the 2D volume form, $\omega$, according to $F =
b(x) \, \omega$. The truncation to these variables is done by
writing
\beq  \label{GP1}
    \exd s^2 = e^{-2\psi(x)} \, g_{\mu\nu}(x) \, \exd x^\mu \exd
    x^\nu + e^{2 \psi(x)} \, \exd s_2^2 \,,
\eeq
where $ds_2^2  = \exd \theta^2 + \lambda^2 \sin^2\theta \, \exd
\varphi^2$ is the line-element on the rugby-ball, whose
defect-angles are related to $\lambda$. Such an ansatz seems to
plausibly capture the relevant physics, and it is much easier to
follow the time-dependence within the truncated action obtained by
following only these modes (and the 4D metric), whose form is
\beq \label{E:4Daction}
    S = c \int d^4x \; \sqrt{-g} \Bigl[ R(g) - (\partial \phi)^2 -
    4 (\partial \psi)^2 - V_\alpha(\phi,\psi) \Bigr] \,,
\eeq
for some constant $c$ and scalar potential $V$. The conclusions
drawn are based on the properties of the solutions to the 4D
equations which follow from this action.\footnote{These arguments
have a historical precedent, since they closely resemble those
made in the 1980's which seemed to indicate 6D solutions with de
Sitter and anti-de Sitter 4-geometries were possible, such as when
the background magnetic flux has monopole number not equal to $\pm
1$. Part of the surprise with the solutions of ref.~\cite{GGP} was
that this expectation was false, since the resulting geometries
turn out to be both squashed and warped, but with flat {\it
intrinsic} 4D geometries.}

The veracity of these conclusions rests on the assertion that the
solutions to the 4D theory of eq.~\eqref{E:4Daction} capture the
same physics as the solutions of the full 6D theory, and there are
two arguments which might (at first sight) indicate that this kind
of reasoning might be true. First, it is known that truncation on
the rugby ball is a {\it consistent} truncation of the 6D field
equations \cite{GibbonsPope}, inasmuch as any solution to the 4D
field equations obtained by truncating the 6D action using the
rugby-ball solution is also automatically a solution of the full
6D field equations. The problem with such a truncation is that it
says nothing about the stability of the solution obtained. For
instance the truncation $X = 0$ is a consistent truncation of the
potential
\beq
    V(X,Y) = A(Y) - B(Y)\, X^2 + C(Y) \, X^4 \,,
\eeq
in precisely the same sense, even if $B(Y) > 0$. However we know
that in this simple case $X = 0$ is a local {\it maximum} of the
potential rather than a minimum, and so the existence of the
truncated solution $X = 0$ provides poor guidance when trying to
infer the endpoint of a time-dependent problem involving both of
the fields $X$ and $Y$. In the 6D example the fields $X$
correspond to the various Kaluza Klein modes, which are typically
nonzero in the known GGP solutions.

The second reason to trust a truncation to 4 dimensions rests on a
firmer footing, since such a truncation is justified as the
leading term in a low-energy expansion provided only that the
truncated variables ({\it i.e.} $X$ in the above example) are
systematically heavy compared with the variables which are kept
({\it i.e.} $Y$ in the above). Since this energy is measured
relative to the local ground state, this criterion does not
mistakenly keep unstable modes as in the above toy example.
Furthermore, in the 6D example, one combination of the two fields
kept in the 4D truncation (the dilaton and radion) is massless,
and so can be understood in such a low-energy theory.
Unfortunately, the other combination, $t = r^2 e^\phi$, is massive
and its mass is known to be precisely the same as the lowest
nontrivial Kaluza Klein states \cite{GibbonsPope,6D4Dsugra}. It
follows that in order to reliably follow the dynamics of the
radion, $\psi$, and the dilaton, $\phi$, {\it separately} in an
effective 4D theory, one must include these Kaluza-Klein states in
the low-energy theory as well, and cannot simply truncate them to
zero by hand, as was done in \cite{GP}.\footnote{See also the
appendix of ref.~\cite{GGPplus} for a discussion of
ref.~\cite{GP}.}

\subsection{Quantum Bulk Contributions: Why There is
Something and Not Nothing}

The previous sections discussed at length the present status of
the classical adjustment of bulk fields to cancel the brane
contributions to the effective 4D cosmological constant. This
section picks up the thread of the main story by returning to the
size of the quantum part of the integration over bulk degrees of
freedom. Given the cancellation between classical bulk physics and
brane contributions, it is this bulk quantum contribution which is
the net nonzero contribution to the observed Dark Energy.

To start: a summary of the result
\cite{Towards,Update,doug,dumitru}. In a nutshell, the upshot of
this discussion is that the effective 4D vacuum energy obtained
within 6D supergravity is similar to what might be expected given
what is known about the vacuum energy within 4D supersymmetric
models. The vacuum energy in a generic 4D supergravity is
typically of order $\Delta V_0 \sim m_{sb}^2 \, M^2$, where
$m_{sb}$ denotes the supersymmetry-breaking scale and $M$ is a
larger scale (often the Planck mass). The suppression of $\Delta
V_0$ by a power of $m_{sb}^2$ is easily understood because
bose-fermi cancellations require the result to vanish in the limit
that $m_{sb} \to 0$ \cite{susycancellation}. For some models the
result can be smaller, with $\Delta V_0 \sim m_{sb}^4$, such as if
special cancellations should ensure the vanishing of the ${\cal
O}(m_{sb}^2)$ term. Essentially the same result holds for 6D
supergravity, with the KK mass splitting, $m_{KK} \sim 1/r$,
playing the role of $m_{sb}$ and the 6D gravity scale, $M_g$,
playing the role of $M$. While for generic 6D supergravities the
4D vacuum energy is of order $M_g^2/r^2$, there are some for which
this leading result cancels to leave a subdominant result which is
of order $1/r^4$. Most interestingly, it is the supergravities
whose 6D field content is that which arises by dimensional
reduction from higher dimensions which appear to enjoy this kind
of cancellation. Now to the details.

\subsubsection{Bulk Loops I: UV Sensitivity}

Since the issue of technical naturalness hinges on the dependence
of the bulk-loop contribution on the heavy (TeV) scale, $M$, the
focus of this first section is on the integration over bulk modes
having energies of order $M$. (The contribution of the lighter
fields is the topic of a later section.) At first sight it is
tempting to think of this calculation as a particularly
complicated 4D calculation which involves an enormous number of 4D
(KK) modes. According to this point of view one would expect each
KK mode to produce the usual 4D $M^4$-dependence in the vacuum
energy, and then expect this result to just get worse once the sum
over all of the KK modes is performed. This turns out to be a
misleading way to calculate, since it hides simplifications which
arise due to important constraints in the 6D theory (like 6D
general covariance and locality) which constrain the contribution
of all modes for which $M \gg m_{KK} \sim 1/r$. These
simplifications are better seen if the vacuum energy is directly
computed within 6 dimensions, which is the point of view followed
here. As usual, you can use any variables you wish to solve a
problem, but if you use the wrong ones you'll be sorry.

In 6 dimensions, if $M \gg 1/r$ then the most $M$-dependent
contributions to the loop-generated vacuum energy can be written
in terms of local interactions in the bulk, and on the branes
\cite{Gilkey}. The brane contributions arise from integration over
bulk fields, because the existence of the branes introduces new
kinds of UV dependence into the bulk \cite{gilkeycones}. This is
how short-wavelength bulk fields `know' about the presence of the
branes, but because of the short wavelength of the modes involved
these effects cannot reach further than a distance $d = {\cal
O}(1/M)$ into the bulk, and so only generate UV-sensitive
effective interactions at the brane positions. The potentially
dangerous contributions to the vacuum energy are any of those
which are proportional to positive powers of $M$, and so it is on
those we focus.

For instance, the six-dimensional effective interactions which can
be generated at one loop depending on positive powers of $M$ in
chiral, gauged 6D supergravity have the form $\Delta S_{\rm eff} =
\Delta S_B + \sum_b \Delta S_b$, where \cite{Update,doug}
\begin{eqnarray} \label{CountertermActions}
    \Delta S_B &=& \int d^6x \; e_6 \, \left[ c_0 M^6 + M^4
    (c_1 \, R + c_1' \,  e^\phi + c_1'' \,(\partial \phi)^2 + \cdots )
    + M^2 ( c_2 \, R^2 + c_2' \, e^\phi \, R + \cdots)
    \right] \nonumber \\
    \Delta S_b &=& \int d^4x \; e_4 \, \left[ M^4(d_0 + d_0' \, e^\phi
    + \cdots) + M^2 (d_1 \, {\cal R} + \cdots)  \right] \,,
\end{eqnarray}
where $R$ is the curvature scalar of the 6D metric, ${\cal R}$ is
the same for the induced 4D metric, and so on. The various
constants, $c_i$, $c_i'$, $d_i$ {\it etc.}, are dimensionless and
the contribution to them due to any one 6D field is of order the
6D loop factor $1/(4 \pi)^3$. For the coefficient of any fixed
power of $M$, the ellipses describe other terms involving the
other fields of 6D supergravity involving the same number of
derivatives of these fields as appear in the explicitly-written
powers of $R$. They also contain terms obtained by replacing
factors of $R \sim 1/r^2$ with factors of\footnote{Keep in mind
here that the classical field equations for chiral, gauged 6D
supergravity imply $e^\phi \sim 1/r^2$ for rugby-ball solutions.}
$e^\phi \sim 1/r^2$, the replacement of $R^2$ by other quadratic
curvature invariants, and so on.

Some of these terms are easily disposed of. First, since the brane
terms are similar to those which were already generated by
integrating out the brane fields, they can be cancelled by the
bulk classical response in precisely the same way.\footnote{It is
possible that the curvature- and dilaton-dependence on the brane
which is generated in this way is responsible for the non-conical
singularities which arise in the GGP solutions.} Next, it is 6D
supersymmetry which disposes of the bulk cosmological constant
(the $M^6$ term), since the contributions of bosons and fermions
explicitly cancel in this term, as they must because local
supersymmetry precludes the existence of a 6D cosmological
constant. Similarly, it happens that the $M^4$ terms also cancel
amongst the elements of a 6D supermultiplet, at least within the
explicit calculations done so far \cite{doug}, for which
supersymmetry breaks in the bulk due to the background gauge flux
not being in the $U(1)_R$ symmetry direction. Even if these $M^4$
terms did not cancel, they would not be dangerous inasmuch as they
are simply renormalizations of the classical 6D action, and such a
renormalization does not affect the classical self-tuning
arguments described above. It does not because the precise value
of the classical couplings (like the 6D Newton constant) is not
important for the cancellation of brane tension and bulk
curvature.

We see from this that the only potentially dangerous terms in
$S_{\rm eff}$ are the $M^2$ terms, and as we shall see these play
a role which is very similar to the $m_{sb}^2 M^2$ terms in 4
dimensions, which are also quadratic in the large mass-scale.
Furthermore, explicit one-loop field-theoretic calculations
\cite{doug} show that the coefficients of these terms do not
vanish in all 6D supersymmetric field theories. However, recent
calculations, which are still in progress, appear to show that 6D
supergravities also exist having a field content for which these
$M^2$ terms also vanish at one loop. For compactifications on
Ricci-flat spacetimes in particular, it appears that the field
content obtained by dimensionally reducing from 10 to 6 dimensions
provides a particularly interesting example for which these $M^2$
terms cancel.

\paragraph{Higher Loops}

Having the $M^2$ terms vanish at one loop is wonderful, but is it
good enough? In particular, what of the contributions of higher
bulk loops? Happily, it is possible to see why these higher loops
need not be dangerous, at least for the gauged, chiral
supergravities \cite{Update}. To see how this works, recall that
in these theories the field equations imply that the background
dilaton is related to the extra-dimensional radius by $e^\phi \sim
1/(Mr)^2$, where the 6D gravity scale $M$ is here made explicit.
But each loop involving bulk particles introduces additional
powers of $e^\phi$, since the dilaton counts loops in 6D
supergravity in the same way that it does in higher-dimensional
supergravities.

So the large-$M$ contributions in 6 dimensions obtained at 2 and
higher loops also involve the same powers of curvature and
derivative as do the 1-loop contributions,
eqs.~\eqref{CountertermActions}, but pre-multiplied by additional
powers of $e^\phi \sim 1/(Mr)^2$. We see that this extra loop
factor is sufficient to tame the overall $M^2$ which
pre-multiplies the curvature squared term, leaving a result which
no longer involves a positive power of $M$. Only the
one-loop-generated curvature-squared terms are dangerous because
of the extremely small size of the bulk coupling, $e^{\phi}$.

One might worry that there may be phenomenological difficulties
with having such tiny bulk couplings, and this would indeed be
true if the brane couplings were also this small since these brane
couplings would include the observed strong and electroweak
interactions. In general, quantities like the brane gauge
couplings (in the Einstein frame) are expected to have an
expansion in powers of $e^\phi$, as does the brane tension in
eq.~\eqref{CountertermActions}, of the form
\beq \label{couplinghierarchy}
    g_b = g_0 + g_1 \, e^\phi + \cdots \,,
\eeq
and the viability of phenomenology on the brane requires $g_0 \ne
0$. This last condition is likely to make it more challenging to
find brane types with the desired properties within string theory.

\subsubsection{Bulk Loops II: Weinberg's Theorem Revisited}

It is worth pausing at this point to revisit the issue of how
these quantum corrections relate to Weinberg's No-Go theorem.
Recall from the discussion of earlier sections that Weinberg's
No-Go theorem states that if a scale invariant theory has a flat
direction, then scale invariance cannot in itself keep this flat
direction from being lifted by quantum effects (even if the scale
invariance has no anomaly, and is a bona fide quantum symmetry).
How does this relate to what has been found about quantum effects
in SLED?

Weinberg's argument is correct when it says that the classical
flat direction is lifted, as is explicitly seen above. The Casimir
energy generates a potential for the classical flat direction,
which can be taken to be parameterized by the extra-dimensional
radius $r$. (Since for SLED the scale invariance is not a quantum
symmetry, these corrections need not be so simple as being quartic
in the appropriate field.) Weinberg's Theorem does {\it not} say
how big these corrections must be, but in a supersymmetric theory
it is natural to expect them to be set by the relevant
supersymmetry-breaking scale.

What is new to the SLED picture is the decoupling it allows
between the supersymmetry-breaking scale of the bulk from that on
the brane \cite{sepsusyscales}. But since the flat direction
involves only bulk fields (like $r$) it is only the supersymmetry
breaking scale in the bulk which is relevant for the lifting of
the flat direction. Crucially, Weinberg provides no argument
against this separation of supersymmetry-breaking scales.

\subsubsection{Bulk Loops III: Massless 6D Modes}

Although quantum bulk contributions to the vacuum energy are
nonzero, they do not generate a cosmological {\it constant}.
Rather, the Casimir energy which is produced describes a scalar
potential for the flat direction of the classical theory.
Understanding this potential is a crucial step towards
understanding the dynamics which explains why the size of the
extra dimensions is as large as we require, $1/r \sim 10^{-2}$ eV.
In this section it is shown what this potential looks like, and
that the mass of this would-be massless field is sufficiently
light as to be cosmologically rolling today. This implies that
within the SLED proposal the dynamics generated for $r$ by the
Casimir energy predicts a time-dependent Dark Energy, with a
particular form for the Dark Energy potential. Furthermore, SLED
automatically explains why the small mass of the radion field is
technically natural {\it in addition} to why the value of the Dark
Energy density is naturally small, using the mechanism first
enunciated in refs.~\cite{ABRS2,ABRS1}.

The total one-loop Casimir energy is obtained by adding the local
contributions, discussed above, of the very massive modes to the
contribution of the lighter modes coming from those fields which
are massless in the 6D sense. The contributions of these lighter
modes have been known for some time for various 6D theories
dimensionally reduced on spheres and torii and in the absence of
branes \cite{casimirtorii,casimirspheres}. More recent
calculations (some of which are in progress) generalize these to
include the branes and their back-reaction, and give similar
results \cite{casimirorbifold,doug,dumitru}.

For instance for the rugby-ball compactifications of chiral,
gauged 6D supergravity $m_{sb} \sim 1/r$ is the
supersymmetry-breaking scale of the bulk sector (in the 4D Jordan
frame), and the resulting vacuum energy is the Casimir energy
obtained by integrating out bulk loops. The result is a potential
for the radion, $V(r)$, which expresses the quantum lifting of the
classical flat direction:\footnote{For toroidal compactifications
of the ungauged 6D supergravities, the potential has a slightly
different form: $V(r,u^i) = W(u^i)/r^4$, where the $u^i$ denote
the various shape moduli and other related scalar fields under
supersymmetry. In particular, no logarithmic terms are generated
in this case at one loop \cite{casimirtorii,dumitru}.}
\beq \label{Vvsr}
    V(r) = \frac{c_2 \, M^2}{r^2} + \frac{c_3}{r^4} \,
    \left[ \log \left( M^2 \, r^2 \right) + C \right] \,.
\eeq
Here $M \sim M_g$ is the TeV scale associated with the integration
over heavy fields, and the dimensionless constants, $c_2$ and
$c_3$, are calculable given the field content of the 6D
supergravity, and are related to the average over the 2 extra
dimensions of those appearing in $S_B$ in
eqs.~\eqref{CountertermActions}. Unlike the logarithmic term, the
constant $C$ depends on the precise matching conditions used at
the higher scale $M$. The logarithmic term arises due to the
renormalization of higher-dimensional effective interactions like
$R^3$, and higher powers of $\log r$ can be expected at
higher-loop orders. The coefficient $c_3$ can be explicitly
computed for a given choice of massless 6D multiplets, and once
integrated over the volume of the 2 extra dimensions is typically
of order $1/(4 \pi)^2$. By contrast, a calculation of the
constants $c_2$ and $C$ depends on the details of the spectrum of
the theory at the electroweak scale, $M$, and of particular
interest are those theories for which this spectrum ensures $c_2 =
0$ \cite{doug}.

In order to understand the dynamics of the model it is necessary
to also include the 4D Einstein and kinetic terms for $r$ in the
low-energy lagrangian \cite{ABRS2}. These lead to the canonical
scalar, $\varphi$, which is related to $r$ by $Mr = e^{k
\varphi/M_p}$, for a positive constant $k$ whose numerical value
is not important for the present purposes. In terms of $\varphi$
the scalar potential, eq.~\eqref{Vvsr}, becomes
\beq
    V(\varphi) = M_p^4 \left[ \hat{c}_2 e^{-2 k \, \varphi/M_p}
    + \hat{c}_3 e^{-4 k \, \varphi/M_p} \left(C + \frac{
    \varphi}{M_p} \right) \right] \,.
\eeq
Here the constants $\hat{c}_i$ are proportional to the $c_i$
appearing in $V(r)$.

Suppose now that we specialize to the present epoch, for which $r
= r_0 \sim 10$ $\mu$m. Then if $c_2 \ne 0$, using $M_p = M^2 r_0$
shows that the Dark Energy density is of order $V \sim M^2/r_0^2$
and the $\varphi$ mass is of order $m_\varphi \sim 1/r_0$. Both of
these are too large to provide an acceptable phenomenology for the
Dark Energy. But the models of interest have $c_2 = 0$, as is
required to ensure a sufficiently small value for $V$: $V \sim
1/r_0^4$. As is clear from the above, this condition ($c_2 = 0$)
automatically also ensures a sufficiently small scalar mass,
$m_\varphi \sim 1/(M_p \, r_0^2) \sim H_0$, to allow a
time-dependent dark energy during the present epoch.

SLED therefore leads to a particular type of time-dependent Dark
Energy having a specific kind of modified exponential potential.
It also predicts that the relevant scalar mass for the late-time
cosmology is of order $m_\varphi^2 \sim V_0/M_p^2$, which allows
the smallness of this mass to be ensured given that the
present-day radius, $r_0$, is large enough to permit $V$ to be
observably small (more about why $r_0$ might be this small below).

\subsubsection{Embedding into More Fundamental Theories?}

Part of the appeal of the SLED proposal is that it involves
supergravity and branes, and so it can plausibly be embedded into
what is at present probably our best-motivated fundamental theory
of physics on the shortest scales: string theory. This section
briefly digresses to summarize what is known about the possibility
of doing so in more detail.

A successful embedding of SLED into string theory consists of
identifying string vacua for which the low-energy excitations
describe both the 6D supergravity of the bulk and the degrees of
freedom trapped on the branes. Indeed, the non-chiral (gauged and
ungauged) 6D supergravities have a known string provenance in this
way as dimensional reductions of 10D string vacua. A complete
string derivation of the chiral 6D theory is not yet known --- see
however \cite{Pedigree} --- although it is plausibly derivable as
the low-energy limit of 10D heterotic string theory compactified
on $K_3$ \cite{susha}.

Knowing how the branes arise within string theory is also
important, and this has several different aspects. It involves
finding string vacua having 3-branes in the effective low-energy
6D supergravity; it involves finding the particles of the Standard
Model living on one of them; and it involves obtaining brane-bulk
couplings which are phenomenologically acceptable
\cite{Towards,susyADS}. A complete understanding of all of these
issues is likely to require an understanding of the origin of the
corresponding 6D bulk supergravity as a precondition. Once the
higher-dimensional origin of the 6D supergravity is known, the
kinds of branes which are available to play the role of the
3-branes in the effective theory may begin to be addressed. Since
compactifications of Type IIA and Type IIB supergravities in 10D
are known to produce the non-chiral, ungauged theories in 6
dimensions, the effective 3-branes in these theories could be
obtained as D3 branes or as higher-dimensional D-branes wrapped
about cycles in the 4 dimensions between 10 and 6. Alternatively,
if the chiral 6D supergravity should emerge as expected as a
compactification of 10D heterotic supergravity on K3, then the
3-branes might instead emerge as NS5-branes wrapped on 2-cycles in
K3.

An important difficulty which any such brane construction must
confront is the large hierarchy between the size of the gauge
couplings on the brane and those in the bulk. That is: why should
$g_0 \ne 0$ in eq.~\eqref{couplinghierarchy}? A discussion of some
of the other issues involved in embedding SLED into a string
framework is given in ref.~\cite{MSLED}.

\subsection{Observational Implications}

Because the SLED proposal modifies physics at sub-eV energy
scales, it has many other observational implications besides its
explanation for the small size of the cosmological constant. This
section very briefly summarizes some of these, including
implications for cosmology, tests of gravity on short and long
distances, and implications for particle physics (including
observable signals at up-coming colliders like the LHC). The
purpose of this section is to summarize some of these other
predictions.

What emerges from this summary is that SLED is unique among Dark
Energy proposals in several ways. First, no other Dark Energy
theory has such wide-ranging implications outside of cosmology,
because no other proposal purports to address the underlying
naturalness issues which require changing physics in the energy
range from $10^{-2}$ eV to several TeV. Second, determining the
size of the extra dimensions requires keeping track of order-one
contributions to the vacuum energy, such as the factors of $2
\pi$. This is a huge improvement over the theoretical
uncertainties of usual Dark Energy proposals, which can easily
involve factors of $10^{50}$. Even more interestingly, the
extra-dimensional sizes to which the Dark Energy leads are close
to the present observational bounds on extra-dimensional sizes.
Best of all, the SLED connection between the Dark Energy and
extra-dimensional size makes it very predictive and falsifiable,
since it is not possible to make the extra dimensions smaller ---
and so easier to hide --- without destroying the explanation of
the observed Dark Energy.

\subsubsection{Bounds on Extra-Dimensional Radii}

This section discusses the most restrictive observational limits
on the possible size the extra dimensions can take. The strongest
such limit comes from the constraint that the energy loss into
Kaluza Klein modes does not provide too efficient an energy-loss
mechanism for supernovae \cite{LEDastrobounds,HR}. This process
has been studied in detail for the special case of the radiation
of gravitons into the bulk, with the result \cite{HR} that an
acceptably small energy-loss rate requires the 6D gravitational
scale, $M_g$, to satisfy $M_g \gsim 8.9$ TeV. For unwarped extra
dimensions this requires the extra-dimensional size to be $r \lsim
10 \; \mu$m.\footnote{Ref.~\cite{HR} quotes $r < 1.6 \; \mu$m, but
uses conventions where the extra-dimensional volume is given by $V
= (2 \pi r)^2$ rather than $V = r^2$.} This limit could be even
stronger for SLED, because there are more states in the bulk into
which energy may be lost. Not all of these may have a nonzero
amplitude for single-particle emission, however, and so the
precise loss rate requires a more detailed calculation. A crude
estimate is
\beq
    \Gamma_{SLED} \approx \Gamma_{LED} {\cal N}\,,
\eeq
where ${\cal N}$ is an estimate of how many more channels are
available in the bulk, relative to the purely gravitational signal
of the nonsupersymmetric case. This leads to the bound
\beq
    M_g > {\cal N}^{1/4}
    8.9 \;\hbox{TeV} \,.
\eeq
In the worst-case scenario where all modes can be radiated we
might have ${\cal N} \sim \frac19 \, \left( {32 + 16 N_g + 8 N_m}
\right)$, where as before $N_g$ and $N_m$ count the number of
gauge and matter multiplets in the bulk. For plausible choices for
${\cal N}$ this leads to a constraint on $r$ which is only
marginally stronger than the LED constraint: $M_g \gsim 12$ TeV
and $r \lsim 5.3 \;\mu$m.

There are also other, nominally stronger, bounds on
extra-dimensional models which come from the non-observance of
Kaluza-Klein modes decaying into photons after having been
produced in supernovae or in the early universe. We ignore these
bounds for the present purposes, since unlike the bound just
discussed they can be completely evaded depending on the details
of the model. For instance, they do not arise if KK modes can
efficiently decay into invisible light modes on other branes. We
regard the model building which such an evasion requires to be
well worth the cost if the resulting theory can make progress on
the much more difficult cosmological constant problem.

\subsubsection{Post-BBN Cosmology}

Besides providing a natural size for the Dark Energy density, the
SLED proposal makes many further predictions concerning the
properties of Dark Energy. First and foremost, it predicts that
the Dark Energy is dynamically evolving in time even now, and so
is not simply a cosmological constant. Furthermore, it accounts
for the amount of Dark Energy {\it and} the very light mass
required by its present-epoch evolution in a way which appears to
be technically natural. Within SLED the evolving 4D scalar field
(or fields) have a microscopic origin as the overall breathing
mode of the 2 large extra dimensions, $r$ (plus possibly various
shape moduli for these same dimensions).

\paragraph{Dark Energy}

Because of this microscopic picture SLED predicts a very specific
form for the Dark Energy field's scalar potential. It must be the
potential of the Casimir energy for the appropriate 6D fields in
the presence of a particular extra-dimensional shape. Because the
extra dimensions are large, the result is robustly a series in
inverse powers of $r$, often with logarithmic corrections, which
is arranged to start at order $1/r^4$. (It is the values of the
constants in this potential which depend on the 6D field content
and the details of the extra-dimensional shape.)

More precisely, within SLED models the Dark Energy arises as a
Casimir energy in the presence of branes within a compact
two-dimensional solution to 6D supergravity. As discussed above,
for the rugby-ball solutions this Casimir energy is given in the
Einstein frame by
\beq \label{VCasimir}
    V(r) = \frac{A}{r^4} \left[1 - a \, \log(M_p r) + \frac{b}{2} \,
    \log^2(M_p r) + \cdots
    \right] + {\cal O}\left( \frac{1}{r^6} \right) \,,
\eeq
where $A$, $a$ and $b$ are dimensionless constants and the
ellipses represent possible terms involving higher powers of $\log
\, r$. Because terms like $\log^\ell r$ first arise at $\ell$-loop
order, within the SLED framework it is natural to expect $b/a \sim
a \sim N_{\rm eff}/(4 \pi)^3$, where $N_{\rm eff}$ provides some
measure of the mismatch between bosons and fermions (and so
depends on the details of how supersymmetry is broken). For
instance, if taken to be of order the total number of bosonic
states in 6D supergravity then $N_{\rm eff} = 16 + 8N_g + 4 N_m$,
where $N_g$ counts the number of gauge multiplets and $N_m$ counts
the number of hypermultiplets. As written, $V$ is positive for all
$r$ provided $A > 0$ and $2b > a^2$.

The potential of eq.~\eqref{VCasimir} is known as an
Albrecht-Skordis potential, and is known to be able to describe a
phenomenologically acceptable description of time-dependent Dark
Energy \cite{AS,ABRS2}. If $a^2 + (b/4)^2 - 2b
> 0$ the potential has a local minimum at $r_-$ and a maximum at
$r_+$, where $4b \log(M_p r_\pm) = b + 4a \pm \Delta$, and $\Delta
= 4[a^2 + (b/4)^2 - 2 b]^{1/2}$. In these cosmologies the
acceleration of the present-day Dark Energy is due to the
evolution of $r$ becoming potential-dominated in the vicinity of
$r_\pm$, and so $r \sim r_\pm$ during the present epoch. Notice
that $r_+$ and $r_-$ can be naturally of order 10 $\mu$m provided
$a/b \approx 70$. Remarkably, such values are not unreasonable for
6D Casimir energies given that for these $a/b \sim (4
\pi)^3/N_{\rm eff}$.

The present-day value of the Dark Energy density in SLED then
becomes
\beq
    \rho \approx V(r_+) = \frac{A}{16 r^4}\Bigl[ b + \Delta \Bigr]
    \,,
\eeq
and so the observed size for $\rho$ is controlled by the values of
$A$, $a$ and $b$. Given estimates for these quantities one can
determine the smallest value for $r$ which is consistent with Dark
Energy observations. Using the reasonable values $A \approx 1$ and
$b \sim a^2$ with $a \sim N_{\rm eff}/(4 \pi)^3$, we see that for
reasonable values of $N_{\rm eff}$, $\rho$ can take the observed
value $(0.003$ eV)${}^4$ with $r$ close to its maximum allowed
value $r \sim 5$ $\mu$m (up to within the uncertainties of the
estimates given). Furthermore, $r$ cannot be much smaller than
this value since this leads to much larger values for $\rho$ due
to the scaling $V \propto 1/r^4$.

Additional constraints also arise within SLED beyond those
normally considered for Albrecht-Skordis potentials because of the
extra-dimensional origin of the scalar fields involved. In
particular, because the overall motion of $r$ implies a change in
the value of Newton's constant over cosmological epochs, it cannot
have changed by more than $\sim 10$\% between the epoch of
nucleosynthesis (BBN) and now --- a condition which can also be
satisfied in these types of models \cite{ABRS2}. Furthermore, the
parameters of the potential must be chosen to ensure that the
present-day potential-energy domination occurs for radii of order
$r \sim 10$ $\mu$m. Such large values are ensured if the
coefficients of the logarithmic terms in the potential are of
order $b/a \sim 1/70$, and remarkably such values arise naturally
within SLED as loop factors.

Although more dedicated studies need to be done, it seems likely
that SLED can share the viable phenomenology of the models
examined in ref.~\cite{ABRS2}. If so, it is likely that the
successful description of present-day cosmology depends on the
particular initial conditions chosen for $r$ just before the BBN
epoch. This would imply that a real understanding of the cosmology
of Dark Energy is likely to require a better theory of its initial
conditions, such as why $r \sim \mu$m should be already true at
the BBN epoch. It must also explain why the various KK modes of
the 2 dimensions should no longer be excited by the time the BBN
epoch is reached, since these also can cause problems for
late-time cosmology (more about this below). What is exciting is
that the possibility of embedding the effective 6D SLED theory
into a more microscopic framework, such as within string theory,
is likely to provide this theory of initial conditions, such as by
explaining them as a consequence of an earlier inflationary epoch
\cite{QuintInflation}.

\paragraph{Dark Matter}

It is not yet known how Dark Matter might fit into the SLED
picture. It would be attractive to be able to use a variant of the
standard Weakly-Interacting Massive Particle (WIMP) paradigm to
describe the Dark Matter as being the relic abundance of a stable
particle having a weak-scale mass and a weak-interaction
annihilation cross section. This paradigm is attractive because
any such particle naturally has the presently-observed abundance
provided only that it was originally in thermal equilibrium with
the other observed particles at epochs when the temperature was of
order their masses. For supersymmetric theories it is common to
use the Lightest Supersymmetric Particle (LSP) as the Dark Matter
WIMP, but this is harder to do for SLED models because the LSP is
the gravitino, which is much too light. Furthermore, ordinary
particles on our brane need not have superpartners at all because
supersymmetry is realized nonlinearly there, removing at a stroke
most of the standard Dark Matter candidates
\cite{nonlinearbranesugra}.

The WIMP proposal is nevertheless very attractive for the SLED
proposal because of the `Why Now?' problem of any model of
time-dependent Dark Energy. This question asks why it should be
that the universal energy density in matter, radiation and Dark
Energy should happen to have been so similar to one another during
the very recent universe. This question is all the sharper given
that these three forms of energy vary very differently as a
function of the universal scale factor as the universe expands. A
clue as to how to understand this problem within the SLED
framework comes from ref.~\cite{AHH}, who argue that the Why Now?
problem has a natural solution if: 1. The Dark Matter is a relic
WIMP abundance, and 2. the Dark Energy is a cosmological constant
which is parametrically of order $\rho \sim (M_w^2/M_p)^4$ (as is
automatically true for the SLED proposal). In such a case, they
argue, their present-day abundances would naturally be close to
one another, since they are both set by ratios of $M_w$ and $M_p$.

Can WIMPS be found within SLED models? The complete answer is not
known, although it is promising that there can an enormous number
of weakly-interacting states available having masses at the weak
scale, the decays of many of which can also be prevented by a
conservation law. As remarked in ref.~\cite{MSLED} there are KK
modes associated with the compactifications from 10 dimensions
down to 6, as well as string modes or winding modes about these
`other' 4 dimensions, any of which could have the required
properties \cite{KKDM}. However in their simplest forms all of
these proposals require the relevant modes to be in thermal
equilibrium at temperatures of order the TeV scale. As is
discussed below, this assumption can be problematic in models
having Large Extra Dimensions.

\subsubsection{Pre-BBN Cosmology}

It happens that in LED and SLED models the real problem with Dark
Matter is not why there is so much of it, but why there is so
little. This is because LED and SLED models can become dangerous
if KK modes in the {\it large} 2 extra dimensions (as opposed to
the smaller 4 weak-scale dimensions) should be stable and be in
equilibrium not much before the BBN epoch. If so, then these modes
can be too abundant and their energy density can contribute too
much to the universal energy budget and either ruin the
predictions of BBN, or make the universe matter-dominated too
early in its history \cite{LEDcosmo}.

This problem needs careful study, but need not be generic because
it depends on fairly model-dependent features of both the geometry
of the extra dimensions and of the universe's thermal history.
When considering LED models it is normally assumed that these
modes were not in equilibrium when ordinary matter had
temperatures above a `normalcy' temperature, $T_*$, which must be
higher than the BBN scale but not as high as the weak scale.
Unfortunately, this option is not available if we wish to explain
the Dark Matter abundance in terms of a thermally generated WIMP,
as in the previous section.

A vulnerable assumption in these bounds is that some Kaluza-Klein
modes are stable because they carry a conserved charge. This is
true in the simplest compactification, such as for torii and
sphere, but is not true when the 2 large dimensions are not
symmetric, or because of the presence of branes  (see for instance
\cite{rabi}). In these cases there need not be conserved charges
in the bulk which the KK modes can carry. If so, then nothings
precludes their quick decay into massless states which would not
dominate the universal energy density. If this were done early
enough it could also avoid ruining the constraints on excess
radiation density at the BBN epoch. It is sufficient to have the
extra dimensions be asymmetric at very early epochs, when the
universe's temperature is above $T_*$, and this need not preclude
their being very symmetric right now.

Indeed, this kind of evolution of the extra dimensions is quite
likely to occur within the SLED proposal because its energy cost
is not high when temperatures are as large as $T_*$. This is
particularly true, given the propensity of the extra dimensions to
become warped and squashed in response to changes in tension on
the various branes. If so, a realistic calculation of the residual
energy tied up in KK modes for the `large' 2 dimensions must await
a better understanding of the time-dependent geometries in 6
dimensions, which is currently under study.

\subsubsection{Tests of Gravity}

There are two sources of deviation from General Relativity which
SLED models predict.

\paragraph{Short-Distance Tests}

First, deviations from Newton's Law are expected on micron length
scales, because these are the scales at which the relevant
effective theory becomes 6-dimensional. The deviations from
Newton's inverse-square law for gravitation must break down in
SLED for the same reason as it does within the earlier LED
proposal: the effects of ordinary graviton exchange begin to
compete at these distances with the exchange of various
Kaluza-Klein modes from the two large extra dimensions. The range
over which the deviations from the inverse-square law arise is set
by the lightest of the nonzero masses of the Kaluza-Klein modes.
Although the precise value of this lightest mass depends on the
details of the precise shape of the extra dimensions, we have seen
above that they are of order $m_{KK} \sim 2\pi \kappa/r$ with
$\kappa = 1$ for a toroidal compactification. It would be
worthwhile computing the form of the predicted force law in more
detail for representative geometries, given the many bulk fields
in SLED whose Kaluza Klein modes could be relevant.

It is central to the SLED proposal that the understanding of the
Dark Energy density fixes $r/(2 \pi)$ to like in the micron range,
and so {\it the absence of deviations from Newton's Law in this
range would rule out the model}. This range is some two orders of
magnitude below the 100 $\mu$m range to which the best present
searches for these effects are sensitive \cite{eotwash}, but are
within reach to the next generation of proposals.

\paragraph{Long-Distance Tests}

Second, the existence of very light 4D scalars (like $r$ or the
any shape moduli) implies that the theory of gravity relevant
right up to the present-day horizon size is a particular type of
4D scalar-tensor model. This follows quite generally as a
consequence of the relaxation mechanism it predicts for the Dark
Energy density, which we have seen leads naturally to one or more
fields, $\varphi$, whose mass is presently incredibly small,
$m_\varphi \sim H_0 \sim 10^{-33}$ eV. Unlike most other Dark
Energy proposals, we have seen how this small scalar mass can be
naturally stable against quantum effects within the SLED
framework.

At the classical level the theory is very predictive because the
breathing mode of an extra dimension is known to couple to
ordinary matter on our brane through their contributions to the
trace of the stress tensor, $T^\mu_\mu$. At first sight this seems
to be a fatal prediction since gravitational-strength couplings of
this form can already be ruled out with some precision. A more
detailed look is again more interesting, and whether these tests
falsify SLED cosmology relies on more detailed calculations of the
$r$-field couplings than are presently available. That they do can
be seen from models such as those of refs.~\cite{ABRS1,ABRS2}, for
which the precise strength and coupling of the light field can
actually depend on some of the details of its cosmological
evolution, since they can evolve during the history of the
universe. They can do so because in the models investigated these
couplings are field-dependent. Since the measurements which
constrain these models are performed only during the present
epoch, they are satisfied provided the couplings happen to be
small at present, and this is what happens in the cosmological
models of \cite{ABRS2}.

Similar effects can arise within SLED scenarios because loop
effects can introduce an $r$-dependence into otherwise
field-independent quantities. Clearly the question of how big
these couplings now are (and whether their motion can be traced
within the domain of perturbation theory) can be addressed in a
more focussed way given a real calculation of the potential and
interactions of the field $\varphi$ (and any other light fields)
from a specific microscopic model for the extra-dimensional
geometry. This provides yet more motivation for exploring the
details of such compactifications, including loop effects, more
closely. We see that, unlike the short-distance tests,
long-distance tests of gravity need not be fatal to SLED models if
deviations from General Relativity on these scales are not seen,
because the size of the present-day couplings of these light
scalars is not central to the explanation of the Dark Energy
properties.

The possibility that there can be very light scalar fields around
providing us with interesting non-standard gravitational physics
on the longest distance scales is particularly exciting given the
recent new opportunities for testing this kind of scenario. It
motivates, in particular, a more careful exploration of the
phenomenology of scalar tensor theories for solar-system tests, as
well as for tests in more exotic settings (like the
recently-discovered system consisting of two pulsars which orbit
one another \cite{HolyGrail}). Given a theoretical framework for
such forces, it becomes possible to make more specific statements
about the nature of their expected couplings, which allows more
focussed analyses of the constraints which can be expected from
these systems.

\subsubsection{Implications for Particle Physics}

Most remarkable of all are the consequences which SLED models have
for particle physics, including most notably the likelihood of
observable missing-energy signals at the Large Hadron Collider
(LHC) which is currently under construction at CERN.

\paragraph{Collider Physics}

A robust consequence of the SLED proposal is the existence of many
bulk Kaluza-Klein modes whose mass spacing is of order $2\pi/r
\sim 0.3$ eV. Although each of these modes is coupled to ordinary
particles only with gravitational strength, their enormous phase
space at TeV energies makes their collective effects enter into
observables at collider experiments with rates of order
\beq
    \Gamma(E) \propto \left( \frac{1}{M^2_p} \right) (E r)^2 \sim
    \left( \frac{E^2}{M_g^4} \right) \,,
\eeq
whose size is controlled by powers of $M_g$ rather than $M_p$. For
this reason the production of missing energy at TeV scale
colliders is a robust signal of the SLED proposal.

Of course, the same arguments apply equally well to the
non-supersymmetric LED proposal, where they again argue for a
robust collider signal, and this has led to extensive studies of
the possible phenomenological signatures of graviton emission into
the bulk \cite{LEDCollider}. Since these studies typically show
that observable effects require $M_g \lsim 1$ TeV, one might worry
that the supernova constraint $M_g \sim 10$ TeV must preclude the
expected SLED collider signal from being observably large.

We argue that this conclusion may be too pessimistic for several
reasons. First, these calculations were done only for the graviton
and we have already seen that radiation rates into the bulk are
more efficient in SLED because of the existence there of all of
the graviton's super-partners \cite{susyaddbounds,SLEDscalars}.
Furthermore, the pessimistic conclusion also relies too heavily on
there being no new physics at energies which are lower than the
scale $M_g$, even though $M_g$ is really only the scale of 6D
gravity as measured by the higher-dimensional Newton constant. As
such, it need not be the threshold at which new physics first
emerges. (A similar error would be made for the weak interactions
if the Fermi constant, $G_F^{-1/2} \sim 300$ GeV, were used to
infer where the scale where the physics of the electro-weak
interactions first appears. In reality we know that this physics
starts at $M_w = 80$ GeV rather than 300 GeV, because of the
appearance of small dimensionless couplings in the relation
between $G_F$ and $M_w$.) In the same fashion, it is likely that
within the SLED proposal the string scale is at or below the scale
$M_g$, making some states potentially available to experiment at
scales of a few TeV.

Furthermore, depending on how it arises within a more microscopic
theory like string theory, there is a possibility in the SLED
picture that there are also Kaluza-Klein and winding states
associated with the `other' 4 compact dimensions obtained when
compactifying from 10 to 6 dimensions. As we saw in earlier
sections, these should also have masses in the TeV regime provided
the 4 small internal dimensions are not strongly warped. In this
regime collider reactions should resemble string collisions near
Planckian energies \cite{StringsPlanck}. A more precise study of
the phenomenology of these modes requires the study of the string
compactifications from 10 to 6 dimensions which lead to the kinds
of 6D models of present interest.

The most likely collider signal (besides the direct production of
new string or KK states) of SLED physics is therefore likely to be
missing energy, which can be produced in association with an
isolated jet or lepton. (For a recent study of bulk-scalar
emission at ATLAS see ref.~\cite{SLEDscalars}.) Such a signal is
unlikely to be confused with the missing energy signals of
alternative proposals, such as the Minimal Supersymmetric Standard
Model (MSSM).

\paragraph{Non-Accelerator Physics}

SLED has tantalizing possibilities for neutrino physics given the
large number of neutral fermions which naturally occur in the
bulk, and which are required by supersymmetry to have vanishing 6D
mass. Furthermore their KK masses are $m_{KK} \sim 1/r \sim
10^{-2}$ eV, which puts them right in the region where neutrino
masses are likely to be. The main obstruction to these models at
present seems to come from astrophysical bounds, since the same
interactions which can provide neutrino masses tend also to
predict an overly abundant rate of energy loss into the extra
dimensions in supernovae \cite{ExtraDimNus}. Study of the
phenomenology of these systems is underway \cite{quim}.

\bigskip

{}From all of the above it is clear that Supersymmetric Large
Extra Dimensions have fascinating implications which seem to cut a
wide swathe across cosmology, astronomy and physics in
laboratories. The proposal relates the Dark Energy in a
fundamental way to observable tests of gravity and to signals in
particle colliders. If true, its verification will be spectacular.
One can only hope!


\begin{theacknowledgments}
  I thank the organizers of the Texas A\&M workshop for their
  hospitality as well as for their kind invitation for me to speak.
  The research described here was done together with numerous
  collaborators to whom I am indebted, including Yashar Aghababaie,
  Andy Albrecht, Georges Azuelos, Hugo Beauchemin, Jim Cline,
  Quim Matias, Susha Parameswaran, Finn Ravndal,
  Constantinos Skordis, Gianmassimo Tasinato, Fernando Quevedo and
  Ivonne Zavala C. I also thank Z. Chacko, G. Gibbons, G. Moore,
  C. Pope and C. Wagner for useful suggestions and criticisms of
  six-dimensional self-tuning. Finally I am obliged to
  Steven Weinberg, both for teaching me how to think about
  effective field theories and for impressing on me the seriousness
  of the cosmological constant problem.
\end{theacknowledgments}

%



\end{document}